\documentclass[journal]{IEEEtran}

\usepackage[T1]{fontenc}
\usepackage{amsmath,amssymb,amsthm,bm,mathtools}
\usepackage{graphicx,booktabs,array,xspace,xcolor}
\usepackage{algorithm}
\usepackage{algpseudocode}
\usepackage{cite}
\usepackage{hyperref}
\newcommand{\secpar}{\lambda}                    % security parameter
\newcommand{\ndim}{N}                            % template dimension
\newcommand{\thr}{\tau}                          % threshold on squared distance
\newcommand{\sfac}{\eta}                         % sampling factor
\newcommand{\nsamp}{t}                           % sample size
\newcommand{\nbad}{w}                            % number of falsified coordinates
\newcommand{\nattempt}{k}                        % number of attempts
\newcommand{\dist}{\delta}                       % squared distance value
\newcommand{\blind}{\rho}                        % commitment blinding
\newcommand{\chal}{\varepsilon}                  % Fiat--Shamir challenge

\newcommand{\vm}[1]{\bm{m}^{\mathrm{#1}}}        % \vm{e}, \vm{p}: enrollment / probe vector
\newcommand{\vr}[1]{\bm{\blind}^{\mathrm{#1}}}   % blinding vector
\newcommand{\vc}[1]{\bm{c}^{\mathrm{#1}}}        % commitment vector: \vc{e},\vc{p},\vc{ee},\vc{pp},\vc{ep}
\newcommand{\mc}[2]{m^{\mathrm{#1}}_{#2}}        % coordinate
\newcommand{\rc}[2]{\blind^{\mathrm{#1}}_{#2}}   % coordinate blinding
\newcommand{\cc}[2]{c^{\mathrm{#1}}_{#2}}        % coordinate commitment

\newcommand{\Gp}{\mathbb{G}}                     % Pedersen group of prime order q
\newcommand{\Zq}{\mathbb{Z}_q}
\newcommand{\Fr}{\mathbb{F}_r}                   % SNARK scalar field
\newcommand{\ord}{q}                             % group order
\newcommand{\fieldord}{r}                        % field order

\newcommand{\Adv}{\mathcal{A}}                   % adversary
\newcommand{\Ext}{\mathcal{X}}                   % extractor
\newcommand{\Sim}{\mathcal{S}}                   % simulator
\newcommand{\Samp}{\mathcal{I}}                  % sampled index set
\newcommand{\Idx}{[\ndim]}                       % all indices
\newcommand{\Chals}{\mathcal{E}}                 % challenge space
\newcommand{\Rel}{\mathcal{R}}                   % matching relation
\newcommand{\Leak}{\mathcal{L}}                  % leakage function

\newcommand{\Setup}{\mathsf{Setup}}
\newcommand{\Enroll}{\mathsf{Enroll}}
\newcommand{\Prove}{\mathsf{Prove}}
\newcommand{\Verify}{\mathsf{Verify}}
\newcommand{\Com}{\mathsf{Com}}                  % Pedersen commitment
\newcommand{\PCom}{\mathsf{PCom}}                % Poseidon commitment
\newcommand{\Hash}{\mathsf{H}}
\newcommand{\Sample}{\mathsf{Sample}}
\newcommand{\Recon}{\mathsf{Recon}}              % distance-commitment reconstruction
\newcommand{\Pass}{\mathsf{Pass}}                % acceptance probability
\newcommand{\AdvSym}{\mathsf{Adv}}

\newcommand{\anc}{\mathsf{anc}}                  % anchor role of a product relation
\newcommand{\bas}{\mathsf{base}}                 % base role of a product relation

\newcommand{\ck}{\mathsf{ck}}                    % commitment key
\newcommand{\crs}{\mathsf{crs}}                  % common reference string
\newcommand{\tx}{\mathsf{tx}}                    % submitted transaction
\newcommand{\lrel}{\ell}                         % relation label in {ee, pp, ep}

\newcommand{\sysname}{\textsc{BioZero}\xspace}
\newcommand{\Full}{\textsc{FullCheck}\xspace}
\newcommand{\Spot}{\textsc{SpotCheck}\xspace}
\newcommand{\Circ}{\textsc{CircuitCheck}\xspace}

\newcommand{\id}{\mathtt{id}}
\newcommand{\nonce}{\mathtt{nonce}}
\newcommand{\ctx}{\mathtt{ctx}}
\newcommand{\pol}{\mathtt{pol}}
\newcommand{\rand}{\mathtt{rand}}

\newcommand{\Ledger}{\mathfrak{L}}

\newcommand{\gas}{\mathrm{gas}}

\newcommand{\sd}[1]{\ensuremath{\|{#1}\|_2^2}}    % squared Euclidean norm
\newcommand{\bino}[2]{\binom{#1}{#2}}

\newtheorem{proposition}{Proposition}
\newtheorem{definition}{Definition}
\theoremstyle{remark}
\newtheorem{remark}{Remark}

\begin{document}

\title{\texorpdfstring{\textsc{BioZero}}{BioZero}: Privacy-Preserving and Publicly Verifiable On-Chain Biometric Authentication via Homomorphic Commitments and Zero-Knowledge Proofs}

\author{Zibin~Lin,
        Taotao~Wang,
        Junhao~Lai,
        Shengli~Zhang,
        Qing~Yang,
        and~Soung~Chang~Liew%
\thanks{Z.~Lin, T.~Wang, J.~Lai, S.~Zhang and Q.~Yang are with the College of Electronics and Information Engineering, Shenzhen University, Shenzhen, China. S.~C.~Liew is with the Department of Information Engineering, The Chinese University of Hong Kong, Hong Kong SAR, China. \textit{Corresponding author: T.~Wang (e-mail: ttwang@szu.edu.cn).}}}

\markboth{}{}
\maketitle

\begin{abstract}
Blockchain-based identity services allow applications across administrative domains to verify biometric authentication results without sharing a trusted biometric server. Such services must protect private templates while limiting both on-chain execution cost and the transaction data required for public verification. We present \sysname{}, a framework for evaluating these requirements under explicit cryptographic assumptions. Exhaustive homomorphic verification (\Full) provides knowledge soundness for distance computed modulo the commitment-group order, but checking every coordinate incurs substantial on-chain cost. At $\ndim{=}128$ on BN254, its benchmark workflow with a threshold stand-in costs $22.1$ million gas. A natural cost-reduction attempt is to sample coordinate checks while retaining this homomorphic structure. This variant (\Spot) lowers benchmark gas by $81.3\%$ to $4.1$ million when seven coordinates are checked, yet still transmits $127{,}844$ bytes of calldata. We prove that changing one product commitment can shift the reconstructed distance to any chosen residue and evade a uniform sample of $\nsamp$ among $\ndim$ coordinates with probability $1-\nsamp/\ndim$. Suppressing this attack to negligible probability therefore requires checking almost every coordinate, removing the scope for substantial savings in coordinate checks. This motivates changing the verification structure: our in-circuit strategy (\Circ) proves commitment openings, coordinate ranges, and the integer distance threshold together. Its complete reference transaction costs $255$ thousand gas with $356$ bytes of calldata, while enforcing the ranges absent from the homomorphic strategies. The client bears the proving cost: at $\ndim{=}128$, median proving time and process peak memory are $0.63$\,s and $822$\,MiB on the evaluated workstation. To assess integer-input suitability, we evaluate the shared biometric front end on two face datasets; every tested width of at least $6$ bits stays within $0.46$ percentage points of float accuracy. These results favor \Circ for the evaluated integer matching task in blockchain-backed cross-domain identity services, provided that committed probes are separately bound to authenticated live captures.
\end{abstract}


\section{Introduction}\label{sec:intro}

Blockchain-based decentralized identity systems can expose an authentication result to applications across administrative domains that do not share a trusted biometric server~\cite{depin,hamer2019private,WANG2025107818}. To use biometrics in this setting, a smart contract must check a match between committed enrollment and probe templates without seeing either vector. The user device submits an authentication transaction, the blockchain executes the verification logic, and relying applications consult the recorded result. The design choice is how much of the match to verify on chain: every coordinate, a sample of coordinates, or one proof of the complete relation. This choice determines the assurance available to applications, the computation executed on chain, and the authentication payload submitted to the network.

Biometric factors can strengthen identity binding when an authenticated acquisition pipeline supplies the probe. Unlike a replaceable token or signing key, a biometric template is difficult to revoke and may reveal sensitive information if exposed. A decentralized authentication layer should therefore keep enrollment and probe templates private while leaving public evidence of the matching decision. The cryptographic proof establishes a relation between committed vectors; the acquisition pipeline must bind the probe to a live capture.

Prior work addresses identity anchoring, credential semantics, and private biometric matching~\cite{yang2020zkdid,sarier2021nontransferable,mao2025zkp,sarier2022privacy,zeng2025ppbiometric}. Exhaustive homomorphic verification links every coordinate to the reconstructed distance, but requires on-chain work and transcript data that grow with the vector dimension. A natural way to reduce the checking work is to sample coordinates while retaining the commitments and distance reconstruction. Reducing the number of checked coordinates addresses only part of this resource demand: a sampled verifier may still require the complete transcript to be submitted and the distance commitment to be reconstructed. The question is whether this change can preserve the exhaustive guarantee, or whether lowering execution and transaction-data costs requires changing the verification structure to a single circuit proof. Answering it also requires distinguishing consistency of modular arithmetic from enforcement of the coordinate ranges needed for integer matching.

% Queue the overview beside its first discussion so it can appear on page 2.
\begin{figure*}[t]
\centering
\includegraphics[width=\textwidth]{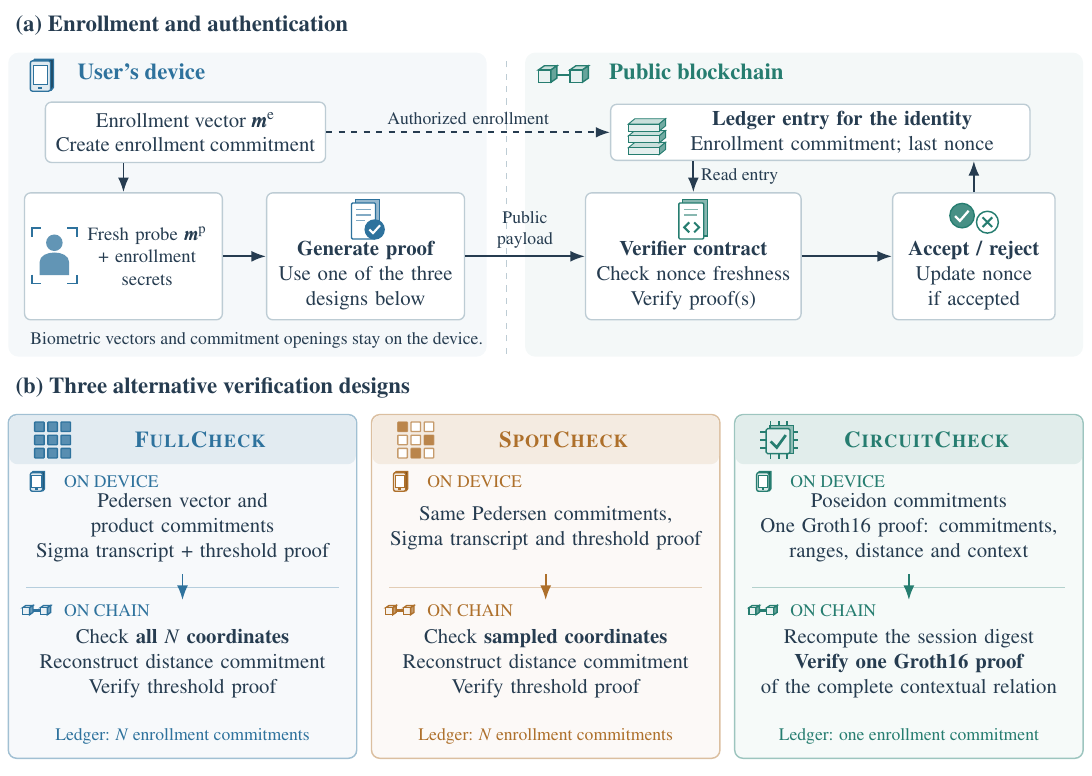}
\caption{Protocol-model overview of the \sysname framework. (a) Authorized enrollment records an identity's commitment on the blockchain. For authentication, the device generates a public payload from the enrollment secrets and a fresh probe; the contract reads the ledger entry, checks freshness and verifies the selected proof path, updating the nonce on acceptance. (b) \Full is the exhaustive baseline, and \Spot is its sampled cost-reduction variant. They share Pedersen commitments, Sigma transcripts and a threshold proof, but check all or sampled coordinates, respectively; both reconstruct the distance commitment. \Circ uses Poseidon commitments and one Groth16 proof of the complete contextual relation, including coordinate ranges.}\label{fig:workflow}
\end{figure*}

We present \sysname{}, a framework for comparing three verification strategies for committed biometric matching under a common enrollment-and-authentication interface. Figure~\ref{fig:workflow} presents this interface and the three alternatives: enrollment records a commitment for identity $\id$, and authentication supplies a probe transcript for verification. In the homomorphic strategies, Pedersen commitments~\cite{pedersen1991} allow the contract to reconstruct a distance commitment; coordinate-wise Sigma arguments link it to the two vectors, and a Groth16 proof~\cite{groth2016size} establishes an opening below the threshold. \Full checks every coordinate and, under the assumptions of Proposition~\ref{prop:full}, provides knowledge soundness for the modular distance relation. We use \Spot, a sampled variant of \Full, to test whether fewer coordinate checks can reduce cost without losing this guarantee. Our single-coordinate attack shows that suppressing the miss probability requires examining nearly all coordinates, leaving little scope for reducing the checking work. \Circ instead changes the commitment and proof structure to establish the complete integer relation, including coordinate ranges, in one circuit. We benchmark transaction costs using a threshold stand-in for the homomorphic strategies.

Kothari et al.~\cite{kothari2026} follow the in-circuit route for fingerprints, with a cancelable minutiae template stored encrypted off chain. We instead compare verification strategies, establish the soundness limitations of sublinear sampling, and instantiate in-circuit verification for fixed-length embedding vectors with coordinate packing and context binding. The owner retains the template locally, while the chain stores a salted commitment; the protocol requires neither a cancelable transform nor a separate off-chain storage service.

The main contributions of this paper are as follows.
\begin{enumerate}
\item \textbf{A common framework and soundness guarantees} (\S\ref{sec:model}, \S\ref{sec:strategies}, \S\ref{sec:security}). We specify enrollment, proof generation, verification, and nonce updates under a common identity and ledger model. Under the stated assumptions, we prove knowledge soundness of \Full for the modular distance relation and of \Circ for the range-constrained integer relation. This distinguishes the exhaustive baseline's guarantee from the additional range constraints required by the biometric task.
\item \textbf{Limits of reducing homomorphic checks by sampling} (\S\ref{sec:security}). We show that one falsified coordinate can move the reconstructed distance to any chosen residue and derive its exact acceptance probability and repeated-attempt amplification. To make the single-coordinate miss probability negligible, sampling must inspect nearly all coordinates. This identifies the limit of reducing verification work through coordinate sampling within the specified homomorphic structure.
\item \textbf{Privacy, freshness, and parameter requirements} (\S\ref{sec:security}). Reusing one response mask across coordinates exposes pairwise coordinate differences; we prove at-most-once nonce acceptance, specify the limits of context binding, and state the group, challenge, and hashing requirements of the homomorphic strategies.
\item \textbf{Implementation evidence and design guidance} (\S\ref{sec:cost}, \S\ref{sec:discussion}). We benchmark complete \Circ transactions and homomorphic workflows with a threshold stand-in. Packed, salted Poseidon commitments and in-circuit range checks give \Circ an authentication cost of approximately $255$ thousand gas at both measured dimensions. At $\ndim=128$, checking seven coordinates reduces stand-in gas by $81.3\%$ relative to \Full, but retains the $127{,}844$-byte transcript and misses a single falsified coordinate with probability $94.5\%$ under uniform sampling. We measure gas and calldata to characterize on-chain execution and submitted transaction-data costs, and report \Circ proving time and memory to characterize the client workload. We also evaluate integer embedding widths on FEI and LFW for the shared biometric front end. The security analysis and on-chain measurements favor \Circ for integer matching in the evaluated setting; the proving and quantization results characterize its resource requirements and input suitability. Appendix~\ref{app:prov} records the measurement provenance.
\end{enumerate}

The remainder of this paper is organized as follows. \hyperref[sec:prelim]{Section~\ref*{sec:prelim}} introduces notation and preliminaries. \hyperref[sec:model]{Section~\ref*{sec:model}} defines the system and threat model, and \hyperref[sec:strategies]{Section~\ref*{sec:strategies}} presents the three verification strategies. \hyperref[sec:security]{Section~\ref*{sec:security}} analyzes their security properties. \hyperref[sec:cost]{Section~\ref*{sec:cost}} reports the implementation, cost, and recognition evaluations. \hyperref[sec:related]{Section~\ref*{sec:related}} reviews related work, \hyperref[sec:discussion]{Section~\ref*{sec:discussion}} discusses the design choice and limitations, and \hyperref[sec:conclusion]{Section~\ref*{sec:conclusion}} concludes. The appendices provide proofs, background, and measurement provenance.

\section{Notation and Preliminaries}\label{sec:prelim}

\subsection{Typographic conventions}
Table~\ref{tab:notation} lists the symbols used throughout the protocol and security analysis. Role tags are upright letters: $\mathrm{e}$ for the enrollment template and $\mathrm{p}$ for the probe; the products of two roles are written $\mathrm{ee}$, $\mathrm{pp}$, and $\mathrm{ep}$.

\begin{table*}[t]
\caption{Notation. One typeface class per kind of object.}\label{tab:notation}
\centering\small
\begin{tabular}{@{}>{\raggedright\arraybackslash}p{3.0cm}>{\raggedright\arraybackslash}p{2.0cm}>{\raggedright\arraybackslash}p{5.3cm}>{\raggedright\arraybackslash}p{6.6cm}@{}}
\toprule
Class & Typeface & Symbols & Meaning \\
\midrule
Scalars, parameters & italic & $\secpar,\ \ndim,\ \thr,\ \sfac,\ \nsamp,\ \nbad,\ \nattempt$ & security parameter, dimension, threshold on $\sd{\cdot}$, sampling factor, sample size, falsified coordinates, attempts \\
Coordinates & italic, indexed & $\mc{e}{i},\ \mc{p}{i},\ \rc{e}{i},\ \dist$ & template coordinates, blinding factors, squared distance \\
Vectors & bold italic & $\vm{e},\ \vm{p},\ \vr{e},\ \vc{e},\ \vc{p},\ \vc{ee},\ \vc{pp},\ \vc{ep}$ & enrollment and probe vectors, blinding vector, commitment vectors \\
Groups, fields & blackboard & $\Gp,\ \Zq,\ \Fr$ & prime-order commitment group, its scalar ring, SNARK scalar field \\
Sets, parties, relations & calligraphic & $\Adv,\ \Ext,\ \Sim,\ \Samp,\ \Chals,\ \Rel,\ \Leak$ & adversary, extractor, simulator, sampled index set, challenge space, relation, leakage \\
Algorithms, functions & sans-serif & $\Setup,\Enroll,\Prove,\Verify,$ $\Com,\PCom,\Hash,\Recon$ & procedures of a strategy, Pedersen and Poseidon commitments, hash, reconstruction \\
Strategy names & small capitals & \Full, \Spot, \Circ & exhaustive, sampled, and in-circuit verification \\
On-chain identifiers & typewriter & $\id,\ \nonce,\ \ctx,\ \pol,\ \rand$ & identity, nonce, context, policy digest, block randomness \\
Ledger state & fraktur & $\Ledger$ & registry of enrollment commitments and last nonces \\
Labels, units & upright & $\mathrm{e},\ \mathrm{p},\ \mathrm{ee},\ \gas$ & role tags, gas \\
\bottomrule
\end{tabular}
\end{table*}

\subsection{Building blocks}
\paragraph*{Smart contract as verifier}
An Ethereum-compatible smart contract serves as the public verifier~\cite{nakamoto2008bitcoin,buterin2014next}. Transaction calldata exposes the submitted transcript on chain; the contract stores enrollment commitments and nonces, checks each transaction, and receives no plaintext templates. Appendix~\ref{app:background} gives further background.

\paragraph*{Pedersen commitments}
Let $\Gp$ be a cyclic group of prime order $\ord$ in which the discrete logarithm problem is hard, and $g,h$ generators with $\log_g h$ unknown. With commitment key $\ck=(\Gp,\ord,g,h)$,
\begin{equation}
\Com_{\ck}(x;\blind)=g^{x}h^{\blind},\qquad x,\blind\in\Zq.\label{eq:pedersen}
\end{equation}
The scheme is perfectly hiding and computationally binding, and it is additively homomorphic: $\Com_{\ck}(x;\blind)\,\Com_{\ck}(x';\blind')=\Com_{\ck}(x{+}x';\blind{+}\blind')$.

\paragraph*{Poseidon commitments}
\Circ commits to a whole vector with a salted, role-separated Poseidon hash~\cite{grassi2021poseidon}, written $\PCom(\vm{};\blind)$, whose arithmetic is native to the SNARK field $\Fr$. Binding assumes collision resistance of the fixed-format hash chain, with role, version, policy, context, dimension, salt, and padded packed coordinates included. The underlying permutation is bijective; the collision assumption concerns its compressed hash outputs. Privacy separately assumes chosen-message hiding with fresh uniform salts and fixed public metadata. Neither collision resistance nor salt entropy alone proves hiding for the concrete Poseidon instance.

\paragraph*{Zero-knowledge proofs}
We use Groth16~\cite{groth2016size} over BN254. A proof consists of two $\Gp_1$ elements and one $\Gp_2$ element; verification costs one pairing check and one scalar multiplication per public input, independent of the circuit size. All three strategies require a Groth16 common reference string $\crs$ and assume a trusted setup.

\paragraph*{Random oracle and Fiat--Shamir}
For the homomorphic strategies, the hash $\Hash$ is modelled as a random oracle. Challenges of their Sigma arguments are derived by Fiat--Shamir~\cite{fiat1986prove} and are drawn from $\Chals=\Zq$.

\paragraph*{Templates and the matching relation}
The intended biometric templates are integer vectors $\vm{}\in\{0,\dots,2^{16}-1\}^{\ndim}$ obtained by quantizing a normalized embedding (\S\ref{sec:quant}). Two templates match when
\begin{equation}
\dist(\vm{e},\vm{p})=\sum_{i=1}^{\ndim}\bigl(\mc{e}{i}-\mc{p}{i}\bigr)^2\le\thr.\label{eq:dist}
\end{equation}
Because $\ndim\,(2^{16}-1)^2<2^{48}$ for $\ndim<2^{16}$, $\dist$ never wraps modulo $\ord$ or $\fieldord$ when both vectors are range-checked. \Circ enforces these coordinate ranges. The homomorphic strategies target a modular distance relation without coordinate-range constraints: \Full establishes this relation under the stated assumptions, whereas \Spot can fail against sparse falsification (\S\ref{sec:security}).

\section{System and Threat Model}\label{sec:model}

\paragraph*{Parties and ledger}
A \emph{prover} (the client device) holds the enrollment secrets and a fresh probe. A \emph{verifier contract} accepts or rejects submitted transactions and updates the ledger on acceptance. The ledger state $\Ledger$ maps each identity $\id$ to its enrollment commitment (a vector $\vc{e}$ for the homomorphic strategies, a single commitment for \Circ) and to the last accepted $\nonce$. The context string $\ctx$ names the domain separator, chain, contract address, group parameters, dimension, and threshold.
The relation-security experiments assume honest enrollment generated independently of the SNARK setup randomness, conditional on the public commitment parameters; subsequent authentication attempts may be adversarial. An external registration authority or ownership policy must authorize the identity-to-commitment assignment, a step omitted from the benchmark contracts. The model excludes malicious registration chosen after observing the CRS and does not establish the sender's ownership of the named identity. Figure~\ref{fig:workflow}(a) illustrates this separation between authorized enrollment and authentication: the verifier reads the enrolled commitment, while a successful authentication updates the nonce.

{
\paragraph*{Relation}
Let $\gamma^{\mathrm{e}}$ and $\gamma^{\mathrm{p}}$ denote the enrollment and probe commitments of a strategy. Commitment openings are interpreted under fixed public parameters. For \Circ, a session instance fixes $(\ndim,\pol,\chi,\thr)$, where $\chi=H_{\Fr}(\ctx,\id,\nonce)$ is the hash-to-field digest; the reference profile additionally requires $0\le\thr<2^{48}$ and $\ndim(2^{16}-1)^2<2^{48}$. Write $X=(\gamma^{\mathrm{e}},\gamma^{\mathrm{p}},\pol,\chi,\thr)$ and $w=(\vm{e},\blind^{\mathrm{e}},\vm{p},\blind^{\mathrm{p}})$. The contextual relation is
\begin{equation}
\begin{aligned}
\Rel_{\thr,\pol,\chi}=\Bigl\{(X;w):\ &\gamma^{\mathrm{e}}=\PCom_{\mathrm{e}}(\vm{e};\blind^{\mathrm{e}};\pol,0,\ndim),\\
&\gamma^{\mathrm{p}}=\PCom_{\mathrm{p}}(\vm{p};\blind^{\mathrm{p}};\pol,\chi,\ndim),\\
&\vm{e},\vm{p}\in[0,2^{16})^{\ndim},\\
&\dist(\vm{e},\vm{p})\le\thr\Bigr\}.
\end{aligned}\label{eq:context-relation}
\end{equation}
The subscripts $\mathrm{e}$ and $\mathrm{p}$ indicate the role tag and common format version. Binding claims assume a fixed verification key and encoding profile, comprising coordinate width, allowed range, packing, padding, and Poseidon parameters; they exclude commitment reuse across profiles. For the homomorphic strategies, $\Rel_{\thr}$ denotes the analogous relation under their fixed commitment parameters and without $(\pol,\chi)$; their natural relation $\Rel^{\mathrm{mod}}_{\thr}$ omits the coordinate range condition and interprets $\dist$ as a residue modulo $\ord$ that must lie in $[0,\thr]$. By contrast, \Circ enforces the range constraints and integer inequality in \eqref{eq:context-relation} (Proposition~\ref{prop:circ}).
}

\paragraph*{Adversary}
The adversary $\Adv$ controls the prover software and the network. It reads all ledger data and every submitted transaction, submits arbitrary transactions, and may adapt each attempt to the outcome of earlier ones. It makes up to $\nattempt$ attempts. For \Spot we model $\rand$ as uniform and fresh per attempt; any bias or predictability of $\rand$ before inclusion can only help $\Adv$.

\paragraph*{Trust assumptions}
Contract execution is correct and the chain is final under the application's confirmation policy. For \Full and \Spot, Pedersen commitments are hiding and binding, and the Fiat--Shamir hash $\Hash$ is a random oracle. For \Circ, the Poseidon binding and hiding assumptions are specified in Proposition~\ref{prop:circ}. We assume Groth16 zero knowledge and adaptive full-state knowledge soundness in the sense of Groth~\cite[Sec.~2.2]{groth2016size}: the extractor has the adversary's state and randomness and obtains a witness for the same output statement. For \Full, this assumption applies to the composed adversary with an internally simulated, lazily sampled oracle and auxiliary ledger and commitment inputs, under a benign distribution of relations and auxiliary inputs and an independently generated CRS. Appendix~\ref{app:proofs} explains the isolated-copy composition. The acquisition pipeline is trusted for capture and feature extraction; liveness and sensor freshness are out of scope. Honest feature extraction produces bounded integers; enforcement against an adversarial prover depends on the verification strategy.

{
\paragraph*{Security goals}
The strategies are compared against six goals, whose attainment is assessed separately for each strategy in \S\ref{sec:security}: \textbf{G1} relation soundness (Definition~\ref{def:sound}); \textbf{G2} offline template privacy (Definition~\ref{def:priv}); \textbf{G3} freshness; \textbf{G4} explicit binding of the public contextual inputs, subject to the limits in \S\ref{sec:security:fresh}; \textbf{G5} public verifiability, because every accepting input and the deployed verifier are observable; and \textbf{G6} minimal leakage for one honestly generated accepting session, as formalized by Definition~\ref{def:priv}. G6 does not cover arbitrary rejected or maliciously formed transactions. Repeated valid threshold queries can reveal information about the template, and the strategies do not bound their cumulative leakage (\S\ref{sec:security:fresh}).
}

{
\begin{definition}[Relation soundness]\label{def:sound}
A strategy $\Pi$ is \emph{relation sound} in the knowledge sense if, for every probabilistic polynomial-time adversary accepted with non-negligible probability, a polynomial-time extractor obtains a relation witness with non-negligible probability for the same accepted public instance. Each quantitative claim below specifies its extraction probability and assumptions; a positive but negligible extraction probability is insufficient. In particular, for \Circ the witness must satisfy $(\gamma^{\mathrm{e}},\gamma^{\mathrm{p}},\pol,\chi,\thr;\vm{e},\blind^{\mathrm{e}},\vm{p},\blind^{\mathrm{p}})\in\Rel_{\thr,\pol,\chi}$ for that transaction's ledger value, probe commitment, policy, and session digest. We abbreviate this fixed-context relation as $\Rel_{\thr}$ where unambiguous.
\end{definition}

\begin{definition}[Offline template privacy]\label{def:priv}
Consider an honest enrollment followed by one honestly generated accepting session. The public view comprises the CRS, enrollment entry, and protocol payload $\tx$ defined in \S\ref{sec:strategies}, excluding the signed Ethereum transport envelope. The leakage is $\Leak=(\mathsf{accept},\ndim,\thr,\pol,\id,\nonce,\ctx)$. In the simulated experiment, a stateful simulator first runs $\mathsf{SimSetup}$ to generate a simulated CRS and retain its simulation state; its transcript phase then takes only $\Leak$ as input and outputs the joint enrollment-and-session view. A strategy $\Pi$ has \emph{offline template privacy} if the real and simulated views are computationally indistinguishable. The definition excludes sender addresses, transaction signatures, fee choices, mempool or block timing, and side channels in local witness handling or proving, and provides no anonymity guarantee for these observations. It also excludes privacy conditioned on a fixed external enrollment record and privacy for arbitrary rejected or maliciously formed payloads.
\end{definition}
}

\paragraph*{Freshness}
A transaction is accepted only if its $\nonce$ exceeds the stored one, and acceptance updates $\Ledger$ atomically. This prevents acceptance twice under the same contract, identity, and nonce. Identity and context are also inputs to the challenge or circuit digest; the limits of this input binding are stated in \S\ref{sec:security:fresh}.

\paragraph*{Non-goals}
Under the assumptions of Propositions~\ref{prop:full} and~\ref{prop:circ}, \Full and \Circ establish extractable knowledge of openings satisfying their respective modular and integer matching relations. Neither result establishes the openings' origin, capture liveness, template uniqueness within a population, or resistance to presentation attacks; capture binding and key custody require separate mechanisms. Kothari et al.~\cite{kothari2026} identify the same capture limitation for their circuit, whose query input is unconstrained.

\section{Three Verification Strategies}\label{sec:strategies}

Figure~\ref{fig:workflow}(b) summarizes the designs and their roles in the comparison. \Full is the exhaustive homomorphic baseline; \Spot varies only which coordinate checks are executed, allowing us to evaluate sampling as a cost-reduction attempt within that structure. \Circ changes the commitments and proof relation to verify the complete integer match in one circuit. The common interface below is instantiated separately for each strategy, with its commitment format, proof relation, and verifier parameters fixed at setup and enrollment. Pedersen and Poseidon enrollment records are not interchangeable.
\begin{itemize}
\item $\Setup(1^{\secpar},\ndim,\thr)\to\mathsf{pp}$: public parameters, including a commitment key or Poseidon parameters and a Groth16 $\crs$.
\item $\Enroll(\mathsf{pp},\id,\vm{e})\to(\gamma^{\mathrm{e}},\mathsf{st})$: $\gamma^{\mathrm{e}}$ is recorded in $\Ledger$; $\mathsf{st}$ stays with the prover.
\item $\Prove(\mathsf{pp},\mathsf{st},\vm{p},\id,\nonce)\to\tx$.
\item $\Verify(\mathsf{pp},\Ledger,\tx)\to\{\mathsf{accept},\mathsf{reject}\}$, updating the stored nonce on acceptance.
\end{itemize}

\subsection{Homomorphic commitments: \Full and \Spot}\label{sec:strategies:hom}

\paragraph*{Commitments}
$\Enroll$ draws $\vr{e}\in\Zq^{\ndim}$ and records $\vc{e}=(\cc{e}{i})_{i}$ with $\cc{e}{i}=\Com_{\ck}(\mc{e}{i};\rc{e}{i})$. $\Prove$ forms the probe commitments $\cc{p}{i}$ and, with fresh blindings, the product commitments
\begin{equation}\label{eq:prodcom}
\begin{aligned}
\cc{ee}{i}&=\Com_{\ck}\bigl((\mc{e}{i})^2;\rc{ee}{i}\bigr),\\
\cc{pp}{i}&=\Com_{\ck}\bigl((\mc{p}{i})^2;\rc{pp}{i}\bigr),\\
\cc{ep}{i}&=\Com_{\ck}\bigl(\mc{e}{i}\mc{p}{i};\rc{ep}{i}\bigr).
\end{aligned}
\end{equation}
By homomorphism, $\Recon$ turns them into a commitment to the squared distance \eqref{eq:dist}:
\begin{equation}
\Recon(\vc{ee},\vc{pp},\vc{ep})=\prod_{i=1}^{\ndim}\cc{ee}{i}\,\cc{pp}{i}\,\bigl(\cc{ep}{i}\bigr)^{-2}=\Com_{\ck}(\dist;\blind_{\dist}),\label{eq:recon}
\end{equation}
The corresponding blinding is
\begin{equation}\label{eq:distance-blinding}
\blind_{\dist}=\sum_{i=1}^{\ndim}(\rc{ee}{i}+\rc{pp}{i}-2\rc{ep}{i}).
\end{equation}

\paragraph*{Per-coordinate argument}
For each relation label $\lrel\in\{\mathrm{ee},\mathrm{pp},\mathrm{ep}\}$, define an \emph{anchor} role and a \emph{base} role by $\anc(\mathrm{ee})=\mathrm{e}$, $\bas(\mathrm{ee})=\mathrm{e}$; $\anc(\mathrm{pp})=\mathrm{p}$, $\bas(\mathrm{pp})=\mathrm{p}$; $\anc(\mathrm{ep})=\mathrm{p}$, $\bas(\mathrm{ep})=\mathrm{e}$. For every coordinate $i$ and label $\lrel$, the prover draws \emph{independent} uniform masks $b^{\lrel}_{1,i},b^{\lrel}_{2,i},b^{\lrel}_{3,i}\in\Zq$ and sends the first messages
\begin{equation}
\alpha^{\lrel}_i=\Com_{\ck}\bigl(b^{\lrel}_{1,i};b^{\lrel}_{2,i}\bigr),\qquad
\beta^{\lrel}_i=\bigl(\cc{\bas(\lrel)}{i}\bigr)^{b^{\lrel}_{1,i}}\,h^{\,b^{\lrel}_{3,i}}.\label{eq:first}
\end{equation}
The challenge binds the whole transcript, including the first messages:
\begin{equation}\label{eq:chal}
\begin{aligned}
\chal=\Hash\bigl(&\ctx\,\|\,\id\,\|\,\nonce\,\|\,\vc{e}\,\|\,\vc{p}\,\|\\
&\vc{ee}\,\|\,\vc{pp}\,\|\,\vc{ep}\,\|\,\bm{\alpha}\,\|\,\bm{\beta}\bigr)\in\Zq.
\end{aligned}
\end{equation}
The responses are
\begin{equation}\label{eq:resp}
\begin{aligned}
z^{\lrel,\mathrm{m}}_i&=b^{\lrel}_{1,i}+\chal\,m^{\anc(\lrel)}_i,\\
z^{\lrel,\mathrm{r}}_i&=b^{\lrel}_{2,i}+\chal\,\blind^{\anc(\lrel)}_i,\\
z^{\lrel,\mathrm{u}}_i&=b^{\lrel}_{3,i}+\chal\bigl(\blind^{\lrel}_i-\blind^{\bas(\lrel)}_i\,m^{\anc(\lrel)}_i\bigr).
\end{aligned}
\end{equation}
The verifier checks, for each coordinate it examines and each label,
\begin{align}
g^{z^{\lrel,\mathrm{m}}_i}h^{z^{\lrel,\mathrm{r}}_i}&=\alpha^{\lrel}_i\bigl(\cc{\anc(\lrel)}{i}\bigr)^{\chal},\label{eq:v1}\\
\bigl(\cc{\bas(\lrel)}{i}\bigr)^{z^{\lrel,\mathrm{m}}_i}h^{z^{\lrel,\mathrm{u}}_i}&=\beta^{\lrel}_i\bigl(\cc{\lrel}{i}\bigr)^{\chal},\label{eq:v2}
\end{align}
that is, six equations per coordinate. Equation~\eqref{eq:v1} shows knowledge of an opening of the anchor commitment; \eqref{eq:v2} shows that $\cc{\lrel}{i}$ commits to the product of the anchored value with the value opened in $\cc{\bas(\lrel)}{i}$.

\paragraph*{Threshold proof}
A Groth16 proof $\pi$ shows that the reconstructed commitment $c'_{\dist}=\Recon(\cdot)$ opens to some $\dist\in[0,\thr]$. By itself this establishes only an opening and a range; the link between $c'_{\dist}$ and the committed vectors comes from the per-coordinate equations. The submitted transaction is
$\tx=(\id,\nonce,\vc{p},\vc{ee},\vc{pp},\vc{ep},\bm{\alpha},\bm{\beta},\bm{z},\pi)$.

\paragraph*{Verification}
Algorithm~\ref{alg:hom} gives $\Verify$. The two strategies differ only in the examined index set $\Samp$.
\begin{itemize}
\item \Full: $\Samp=\Idx$.
\item \Spot: $\nsamp=\min\{\ndim,\lceil\sfac\log_2\ndim\rceil\}$ and $\Samp$ consists of the first $\nsamp$ distinct values of $1+(\Hash(\mathsf{seed}\|j)\bmod\ndim)$, where $\mathsf{seed}=\Hash(\chal\|\id\|\nonce\|\rand)$.
\end{itemize}

\begin{algorithm}[t]
\caption{$\Verify$ for \Full and \Spot}\label{alg:hom}
\begin{algorithmic}[1]
\Require $\tx$, ledger $\Ledger$, parameters $\mathsf{pp}$, sampling factor $\sfac$ (\Spot only)
\If{$\nonce\le\Ledger[\id].\nonce$} \State \Return $\mathsf{reject}$ \EndIf
\State load $\vc{e}$ from $\Ledger[\id]$; compute $\chal$ by \eqref{eq:chal}
\State $\Samp\gets\Idx$ \textbf{(\Full)}\quad or\quad $\Samp\gets\Sample(\chal,\id,\nonce,\rand,\ndim,\nsamp)$ \textbf{(\Spot)}
\For{$i\in\Samp$, $\lrel\in\{\mathrm{ee},\mathrm{pp},\mathrm{ep}\}$}
  \If{\eqref{eq:v1} or \eqref{eq:v2} fails} \State \Return $\mathsf{reject}$ \EndIf
\EndFor
\State $c'_{\dist}\gets\Recon(\vc{ee},\vc{pp},\vc{ep})$ \Comment{always over all $\ndim$ coordinates}
\If{Groth16.Verify$(\mathsf{vk},(c'_{\dist},\thr),\pi)=0$} \State \Return $\mathsf{reject}$ \EndIf
\State $\Ledger[\id].\nonce\gets\nonce$; \Return $\mathsf{accept}$
\end{algorithmic}
\end{algorithm}

\subsection{One proof over the whole relation: \Circ}\label{sec:strategies:circ}

\Circ places the commitment-opening, coordinate-range, and distance checks in one Groth16 circuit over $\Fr$. The contract separately checks nonce freshness and recomputes the session digest supplied as a public input. Let $\chi=\Hash(\ctx\|\id\|\nonce)$ reduced into $\Fr$ be the per-transaction digest. $\Enroll$ draws a salt $\blind^{\mathrm{e}}\in\Fr$ and records the single commitment $\gamma^{\mathrm{e}}=\PCom(\vm{e};\blind^{\mathrm{e}};\pol)$. $\Prove$ draws $\blind^{\mathrm{p}}$, forms $\gamma^{\mathrm{p}}=\PCom(\vm{p};\blind^{\mathrm{p}};\chi)$, and proves, with public inputs $(\gamma^{\mathrm{e}},\gamma^{\mathrm{p}},\pol,\chi,\thr)$ and witness $(\vm{e},\blind^{\mathrm{e}},\vm{p},\blind^{\mathrm{p}})$, that:
\begin{enumerate}
\item every coordinate of $\vm{e}$ and $\vm{p}$ lies in $[0,2^{16})$;
\item $\gamma^{\mathrm{e}}$ and $\gamma^{\mathrm{p}}$ are the salted, role-separated commitments of the two vectors, the first absorbing $\pol$ and the second absorbing $\chi$;
\item the distance in \eqref{eq:dist} is computed in $\Fr$, fits in $48$ bits, and satisfies $\dist\le\thr$.
\end{enumerate}
The reference circuit packs fifteen 16-bit coordinates into each field element. At $\ndim{=}128$ and $\ndim{=}512$, respectively, this reduces the constraint counts from $38{,}464$ and $151{,}360$ for reconstructed one-coordinate-per-element circuits to $7{,}660$ and $27{,}118$, reductions of $80.1\%$ and $82.1\%$. Appendix~\ref{app:prov} describes the reconstructed baselines. $\Verify$ checks the nonce, loads $\gamma^{\mathrm{e}}$, recomputes $\chi$, runs the Groth16 verifier on the five public inputs, and updates the nonce. Verification requires one pairing check and five public-input multiplications, independent of $\ndim$.

\subsection{Comparison of the strategies}
Table~\ref{tab:strategies} summarizes what each strategy verifies. \Full is the exhaustive baseline: it checks all six equations at every coordinate and reconstructs the distance commitment. \Spot changes only the number of examined coordinate equations, isolating the saving from sampling. Both homomorphic strategies retain $\ndim$ enrollment commitments in the model, the full submitted transcript, and the $O(\ndim)$ reconstruction \eqref{eq:recon}. \Circ instead uses a vector commitment and one proof of the integer relation, with five public inputs independent of dimension. Section~\ref{sec:security} establishes which guarantees these verification choices provide.

\begin{table*}[t]
\caption{What each strategy verifies on chain. The final column summarizes the assumptions or limitations of relation soundness (Definition~\ref{def:sound}).}\label{tab:strategies}
\centering\small
\begin{tabular}{@{}>{\raggedright\arraybackslash}p{2.4cm}>{\raggedright\arraybackslash}p{3.1cm}>{\raggedright\arraybackslash}p{2.3cm}>{\raggedright\arraybackslash}p{3.1cm}>{\raggedright\arraybackslash}p{1.9cm}>{\raggedright\arraybackslash}p{3.1cm}@{}}
\toprule
Strategy & Per-coordinate equations examined & Enrollment storage & What the SNARK proves & Coordinate range enforced & Soundness assumptions or limitations \\
\midrule
\Full & all $\ndim$ (six each), plus $\Recon$ & $\ndim$ group elements & $c'_{\dist}$ opens to a value in $[0,\thr]$ & no & Pedersen binding, Fiat--Shamir, full-state Groth16 extraction (Prop.~\ref{prop:full}) \\
\Spot & $\lceil\sfac\log_2\ndim\rceil$ indices, capped at $\ndim$ (six each), plus $\Recon$ & $\ndim$ group elements & as \Full & no & sparse inconsistency passes with the probability in Proposition~\ref{prop:accept} \\
\Circ & none outside the circuit & one commitment & the whole relation $\Rel_{\thr}$, including range and commitments & yes ($16$ bits) & Groth16, collision resistance of Poseidon \\
\bottomrule
\end{tabular}
\end{table*}

\section{Security Analysis}\label{sec:security}

The comparison separates two requirements: linking the distance to every committed coordinate and enforcing integer coordinate ranges. \Full provides the first under the extraction and binding assumptions below; \Spot can lose it through sparse falsification; \Circ enforces both in its proved relation. We first determine whether sampling can preserve the exhaustive guarantee while avoiding most coordinate checks, then state the positive guarantees for exhaustive and in-circuit verification.

For the homomorphic strategies, a transcript is \emph{falsified} in a set $W\subseteq\Idx$ of $\nbad=|W|$ coordinates if, for $i\in W$, at least one of the six equations \eqref{eq:v1}--\eqref{eq:v2} fails. Proofs are in Appendix~\ref{app:proofs}.
The results assume honest enrollment and the group, setup, and randomness conditions of \S\ref{sec:model}. For \Full, the threshold circuit must prove the Pedersen opening; the benchmark circuit in \S\ref{sec:cost} omits this constraint. Exact sampling probabilities use an independent uniform subset, and the $1/|\Chals|$ terms use an ideal uniform field-valued random oracle.

\subsection{Sampling does not give negligible soundness error}\label{sec:security:sampling}

\begin{proposition}[Acceptance of falsified transcripts]\label{prop:accept}
Let $\Samp$ be a uniformly random set of $\nsamp$ distinct indices, independent of $W$, with $0\le\nsamp\le\ndim$ and $0\le\nbad\le\ndim$. A transcript falsified in $W$, and otherwise valid, passes the per-coordinate checks with probability
\begin{equation}\label{eq:sampling-pass}
\Pass(\ndim,\nsamp,\nbad)=\frac{\bino{\ndim-\nbad}{\nsamp}}{\bino{\ndim}{\nsamp}}\ \le\ \Bigl(1-\frac{\nbad}{\ndim}\Bigr)^{\nsamp},
\end{equation}
and $\Pass(\ndim,\nsamp,1)=1-\nsamp/\ndim$ exactly. In particular, a soundness error of at most $\varepsilon$ against a single falsified coordinate requires $\nsamp\ge(1-\varepsilon)\ndim$. With $\nsamp=\lceil\sfac\log_2\ndim\rceil$ the error tends to $1$ as $\ndim$ grows, for every fixed $\sfac$.
\end{proposition}

\begin{table}[t]
\caption{Probability that a transcript falsified in $\nbad$ of $\ndim{=}128$ coordinates passes the per-coordinate checks (Proposition~\ref{prop:accept}; computed in closed form).}\label{tab:accept}
\centering\footnotesize
\begin{tabular}{lrrrrr}
\toprule
Strategy & $t$ & $w{=}1$ & $w{=}2$ & $w{=}6$ & $w{=}13$ \\
\midrule
\textsc{SpotCheck}, $\eta{=}1$ & 7 & 94.53\% & 89.32\% & 70.86\% & 46.35\% \\
\textsc{SpotCheck}, $\eta{=}2$ & 14 & 89.06\% & 79.24\% & 49.17\% & 20.47\% \\
\textsc{SpotCheck}, $\eta{=}4$ & 28 & 78.12\% & 60.90\% & 21.98\% & 3.36\% \\
\textsc{SpotCheck}, $\eta{=}8$ & 56 & 56.25\% & 31.45\% & 2.88\% & 0.03\% \\
\textsc{FullCheck} & 128 & 0.00\% & 0.00\% & 0.00\% & 0.00\% \\
\bottomrule
\end{tabular}

\end{table}

Table~\ref{tab:accept} reports the acceptance probabilities at $\ndim{=}128$. With $\sfac{=}1$ ($\nsamp{=}7$), one falsified coordinate passes with probability $94.5\%$; even with $\sfac{=}8$ ($\nsamp{=}56$) it passes with $56.3\%$.

When many coordinates are falsified, the bound is useful: if $\nbad\ge\rho\ndim$ for some $0<\rho<1$, then $\Pass\le(1-\rho)^{\nsamp}$, so $\nsamp\ge\ln(1/\delta)/(-\ln(1-\rho))$ samples reduce the acceptance probability to at most $\delta$. Sampling therefore detects \emph{dense} falsification. Under the conditions of Proposition~\ref{prop:single}, however, one falsified coordinate suffices.

\begin{proposition}[One falsified coordinate suffices]\label{prop:single}
{Let $0\le\thr<\ord$. Let a prover know an opening $(\vm{e},\vr{e})$ of $\vc{e}$ and a probe $\vm{p}$ whose distance residue $\dist(\vm{e},\vm{p})\bmod\ord$ exceeds $\thr$. For every target $\dist^{\ast}\in[0,\thr]$ and coordinate $j$, it can replace only product commitment $\cc{ep}{j}$, and choose responses so that exactly coordinate $j$ fails the consistency checks, while knowing a valid opening of the reconstructed commitment to $\dist^{\ast}$. Under the independent uniform-subset sampling model, acceptance occurs exactly when $j\notin\Samp$, with probability $1-\nsamp/\ndim$. For bounded templates with $\ndim(2^{16}-1)^2<\ord$, the distance residue equals the integer distance.}
\end{proposition}

Together, Propositions~\ref{prop:accept} and~\ref{prop:single} give a necessary condition for the proposed saving: limiting the single-coordinate attack's acceptance probability to $\varepsilon$ requires $\nsamp\ge(1-\varepsilon)\ndim$. At most an $\varepsilon$ fraction of coordinates can then be left unchecked, so negligible error removes the scope for a substantial reduction in coordinate checks. The threshold proof certifies only that $c'_{\dist}$ opens to a value in $[0,\thr]$ and does not prevent this attack. The BN254 consistency-check benchmarks illustrate the predicted sample behavior: each sampled configuration accepted a transcript with one falsified, unsampled coordinate, whereas the exhaustive check rejected it (Appendix~\ref{app:prov}).

\begin{remark}
{The knowledge-soundness counterexample requires an efficiently constructible above-threshold distance residue and a non-negligible miss probability. Any extractor producing a matching witness for the same enrollment and probe commitments would have to change at least one known message opening, breaking Pedersen binding. When $\nsamp=\ndim$, the miss probability is zero and \Spot coincides with \Full. When $\thr=\ord-1$, the modular distance condition is vacuous and this counterexample does not apply. For bounded integer templates, no above-threshold pair exists if $\thr\ge\ndim(2^{16}-1)^2$.}
\end{remark}

\begin{proposition}[Repeated attempts]\label{prop:attempts}
{Suppose that, conditional on every prior history, the current falsified set has fixed size $\nbad$ and is fixed before a fresh uniform sample is revealed, and all other checks pass. The probability of at least one acceptance in $\nattempt$ attempts is $1-(1-\Pass(\ndim,\nsamp,\nbad))^{\nattempt}$. With predictable sampling and $1\le\nbad\le\ndim-\nsamp$, an otherwise honest prover can instead corrupt only responses at unsampled coordinates and pass with certainty; this is a consistency-predicate counterexample, not by itself a false-distance attack. For the distance-shift attack of Proposition~\ref{prop:single}, fresh independent offline candidates give success $1-(1-P)^K$ after $K$ trials, where $P=1-\nsamp/\ndim$.}
\end{proposition}

\begin{table}[t]
\caption{Probability of at least one acceptance within $\nattempt$ attempts at $\ndim{=}128$ (Proposition~\ref{prop:attempts}).}\label{tab:attempts}
\centering\footnotesize
\begin{tabular}{lrrrr}
\toprule
Strategy & $k{=}1$ & $k{=}2$ & $k{=}5$ & $k{=}10$ \\
\midrule
\textsc{SpotCheck}, $\eta{=}8$, $w{=}1$ & 56.2\% & 80.9\% & 98.4\% & 100.0\% \\
\textsc{SpotCheck}, $\eta{=}8$, $w{=}6$ & 2.9\% & 5.7\% & 13.6\% & 25.3\% \\
\textsc{SpotCheck}, $\eta{=}4$, $w{=}6$ & 22.0\% & 39.1\% & 71.1\% & 91.6\% \\
\textsc{SpotCheck}, $\eta{=}1$, $w{=}6$ & 70.9\% & 91.5\% & 99.8\% & 100.0\% \\
\bottomrule
\end{tabular}

\end{table}

Table~\ref{tab:attempts} shows the effect: at $\sfac{=}8$ and one falsified coordinate, five attempts succeed with probability $98.4\%$. On-chain attempt limits require persistent records of rejected transactions. They do not prevent offline search when $\rand$ is predictable, since an adversary can run $\Verify$ locally and submit only passing transcripts.

\subsection{Exhaustive and in-circuit strategies}\label{sec:security:full}

\begin{proposition}[\Full]\label{prop:full}
{Let $0\le\thr<\ord$ and let $\Adv$ make at most $q_{\Hash}$ random-oracle queries and induce acceptance with probability $p$. Set $Q=q_{\Hash}+1$, including a possible final verifier query. Assume the full-state Groth16 knowledge soundness specified in \S\ref{sec:model} for the composed prover and oracle wrapper. Let $\delta_{\mathsf{G16}}$ bound its accepting-but-no-valid-threshold-witness probability, and let $\delta_{\mathsf{bind}}$ bound any conflicting-message opening found in the joint experiment. Then a polynomial-time extractor outputs a witness for the accepted transaction's $\Rel^{\mathrm{mod}}_{\thr}$ with probability at least
\begin{equation}\label{eq:full-extraction-bound}
\max\!\left\{0,\ p\Bigl(\frac{p}{Q}-\frac{1}{|\Chals|}\Bigr)-\delta_{\mathsf{bind}}-\delta_{\mathsf{G16}}\right\}.
\end{equation}
For polynomial $Q$ and negligible bad-event probabilities, non-negligible acceptance implies non-negligible extraction. Appendix~\ref{app:proofs} composes the full-state extractor with the fork by copying the first execution. The result concerns field-valued coordinates without $[0,2^{16})$ range constraints.

Separately, under Groth16 zero knowledge, \Full has the honest accepting-session privacy of Definition~\ref{def:priv}, with advantage at most $\AdvSym^{\mathsf{zk}}+q_{\Hash}/\ord$, where here $q_{\Hash}$ bounds the observer's oracle queries.}
\end{proposition}

\begin{remark}[Range]\label{rem:range}
For the BN254 scalar order $\ord\equiv1\pmod4$ and $\ndim\ge2$, choose $s\in\Zq$ with $s^2=-1$, enroll $\vm{e}=\bm0$, and use $\vm{p}=(-1,-s,0,\ldots,0)$ as field residues. The modular squared distance is $1+s^2=0$, so honest product equations and a zero-distance opening proof pass for every $\thr\ge0$. Yet the first probe coordinate is $\ord-1$, outside the template range. Enforcing integer-template semantics therefore requires additional range arguments.
\end{remark}

\begin{proposition}[\Circ]\label{prop:circ}
{Assume Groth16 adaptive knowledge soundness and collision resistance of the fixed-format $\PCom$ family. Fix $1\le\ndim<2^{16}$, $0\le\thr<2^{48}$, canonical public field elements, and the policy and context of Definition~\ref{def:sound}. (i)~For an adversary accepted with probability $p$, its knowledge extractor outputs a witness for the same public instance of $\Rel_{\thr}$ with probability at least $p-\AdvSym^{\mathsf{G16}}$, where $\AdvSym^{\mathsf{G16}}$ is the probability of acceptance without a valid extracted witness in the knowledge experiment. (ii)~For honestly generated enrollment, the extracted vector agrees with the enrolled vector except with probability $\AdvSym^{\mathsf{cr}}$ of a commitment collision. (iii)~Range checks and $\ndim(2^{16}-1)^2<2^{48}<\fieldord$ make the distance comparison an integer comparison. (iv)~Additionally, assume that independently salted commitments hide either of two equal-length messages at fixed public metadata, with standard message-hiding advantage $\AdvSym^{\mathsf{hide}}$, even given independently generated simulated-CRS auxiliary information. Then the honest accepting-session privacy advantage is at most $\AdvSym^{\mathsf{zk}}+2\AdvSym^{\mathsf{hide}}$.}
\end{proposition}

\subsection{Privacy of homomorphic auxiliary responses}\label{sec:security:aux}

\begin{proposition}[Shared masks leak coordinate differences]\label{prop:leak}
Suppose one mask $b$ is used for the responses $z_i=b+\chal\,m_i$ of all coordinates. For $\chal\ne0$, any observer computes $m_i-m_j=(z_i-z_j)\,\chal^{-1}$ for all $i,j$. For an honest transcript with uniform challenge in $\Zq$, the excluded $\chal=0$ event has probability $1/\ord$. With independent uniform masks and a fixed challenge independent of them, the response vector is uniform and its differences reveal no template differences. For honest Fiat--Shamir generation in the ROM, this argument applies to a fresh first-message query; prior-query exceptions are covered by Proposition~\ref{prop:full}. Response marginals alone do not establish privacy of the complete transcript.
\end{proposition}

On synthetic vectors with entries in $[0,1000)$ and $\ndim{=}128$, the difference attack recovered all pairwise differences in $1000$ of $1000$ trials, and the coordinate range alone left a median of $14$ candidates for the first coordinate; with independent masks it recovered only the trivial difference $m_1-m_1$ ($1/128$ of the differences relative to the first coordinate). The specification of \S\ref{sec:strategies:hom} therefore draws independent masks for every session, coordinate, and relation. The same remark applies to the blinding responses $z^{\lrel,\mathrm{r}}_i$ and $z^{\lrel,\mathrm{u}}_i$.

\subsection{Nonce freshness, context binding, and outcome leakage}\label{sec:security:fresh}

\begin{proposition}[At-most-once acceptance]\label{prop:fresh}
For a fixed contract and identity, with persistent enrollment and atomic monotone nonce updates, a transaction with nonce $\nonce$ cannot be accepted again after that nonce has been accepted.
\end{proposition}

{\emph{Argument.} Acceptance records $\nonce$, and later acceptance requires a strictly larger nonce. This state transition gives at-most-once acceptance independently of hash assumptions. Context is also included in the homomorphic challenge and, for \Circ, in the public input $\chi=H_{\Fr}(\ctx,\id,\nonce)$. The implementation uses Keccak followed by reduction modulo $\fieldord$; collision resistance of this hash-to-field map is assumed directly, since a collision after reduction need not be a Keccak collision. Distinct contexts yield distinct $\chi$ values except with the collision probability of $H_{\Fr}$. This binds the public inputs but does not establish transcript non-malleability or prevent front-running, since the contracts do not bind the caller or intended application action.}

Public verifiability (G5) follows from observable inputs and verifier code. G6 concerns the honest accepting-session view in Definition~\ref{def:priv}. Rejection may have several causes and does not necessarily reveal a negative distance comparison, while repeated valid threshold comparisons can disclose more than one bit overall. Nonces remain unchanged on failure, so they do not limit failed attempts. Privacy for arbitrary rejected transactions and adaptive online comparisons requires a separate analysis.

\subsection{Parameter requirements}\label{sec:security:params}

\begin{table*}[t]
\caption{Parameter requirements of the homomorphic strategies, the setting used in this paper, and the consequence of violating each requirement.}\label{tab:params}
\centering\small
\begin{tabular}{@{}>{\raggedright\arraybackslash}p{2.4cm}>{\raggedright\arraybackslash}p{4.2cm}>{\raggedright\arraybackslash}p{3.3cm}>{\raggedright\arraybackslash}p{6.0cm}@{}}
\toprule
Item & Requirement & This paper & If violated \\
\midrule
Commitment group & hard discrete log; prime order $\ord\ge2^{2\secpar}$ is a necessary generic-attack size condition, not sufficient by itself; $\log_g h$ unknown & BN254 $\Gp_1$; benchmark key $h=5g$ is not binding & with $p\approx2^{26}$, $\log_g h$ is found in about $\sqrt{p}$ group operations and any commitment can be opened to any value \\
Challenge space & $|\Chals|\ge2^{\secpar}$ & $\Chals=\Zq$ & a cheating prover guesses the challenge with probability $1/|\Chals|$ per attempt \\
Masks and blindings & uniform in $\Zq$, independent per session, coordinate, relation & independent and uniform in $\Zq$ & shared or low-entropy masks expose coordinate differences (Proposition~\ref{prop:leak}) or the committed values \\
Challenge input & $\ctx$, $\id$, $\nonce$, all commitments, all first messages & all of them \eqref{eq:chal} & the challenge no longer depends on the first messages, and the Fiat--Shamir argument does not apply \\
Sample set & derived from unpredictable $\rand$ & modelled as uniform and fresh & predictability enables response-only evasion or offline grinding of distance-shift candidates (Proposition~\ref{prop:attempts}) \\
\bottomrule
\end{tabular}
\end{table*}

Table~\ref{tab:params} summarizes the parameter requirements. An undersized group allows commitment forgery: for a modulus of about $2^{26}$ (we checked $p=67{,}108{,}859$), baby-step giant-step finds $s=\log_g h$ with $5{,}793$ baby steps; given $s$, a commitment $g^{m}h^{\blind}$ can be opened to any other value $m'$ with $\blind'=\blind+(m-m')s^{-1}$, so the commitments are not binding. Cost figures obtained with such a group describe a configuration without cryptographic security. Section~\ref{sec:cost} therefore measures the homomorphic checks at a cryptographically sized group.

\section{Implementation and Evaluation}\label{sec:cost}

The evaluation measures the cost of the exhaustive homomorphic baseline, the savings and residual costs of sampling within that structure, and the resources required by the complete circuit relation. Transaction gas and calldata characterize the verification choices, while proving time and memory characterize the \Circ client implementation. The quantization experiments separately assess the shared integer inputs; they do not rank the verification strategies.

\subsection{Method and provenance}
{All current experiments were executed on \texttt{enine} (Ubuntu~22.04.5/Linux~6.8.0-138 x86-64, Intel Xeon w7-3445, 251.36\,GiB RAM). The \Spot and \Full measurements use Node.js~22.22.3, \texttt{solc}~0.8.36, and Ganache~7.9.2 (chain ID 31337, Shanghai rules). The \Circ complete-transaction and packed-circuit measurements additionally use Circom~2.2.3 and snarkjs~0.7.6. Each deployed verifier matches its circuit's verification key. Appendix~\ref{app:prov} records the source files, hashes, setup provenance, and exact environments.}
\begin{itemize}
\item \textbf{Homomorphic stand-in workflow (\Full, \Spot).} A BN254 contract measures the transaction workflow of Algorithm~\ref{alg:hom}, with the threshold stand-in described below. It checks and updates the nonce, checks the submitted $\vc{e}$ against a digest stored at enrollment, computes $\chal$ by \eqref{eq:chal} over all commitments and first messages, derives $\Samp$ from \texttt{block.prevrandao} (\Spot), evaluates the six equations for every examined coordinate, performs the reconstruction \eqref{eq:recon} over all $\ndim{=}128$ coordinates, and calls a Groth16 verifier for the threshold statement. Because $\Samp$ is fixed only when the transaction is included, the prover submits the whole transcript and the contract examines $\nsamp$ coordinates of it. {For every reported setting, three authentications with fresh transcripts were accepted and the median measured transaction gas is reported. At every \Spot sweep label, replay, a threshold proof reused under a new nonce, a substituted enrollment vector, and a modified response were rejected; the \Full campaign exercised the same required controls.}
\item \textbf{Threshold verifier.} The threshold statement has five public inputs: $c'_{\dist}$ as four $128$-bit limbs, and $\thr$. The measured circuit range-checks $\dist$ and enforces $\dist\le\thr$, but it does not prove that $c'_{\dist}$ opens to $\dist$, which would require non-native BN254 arithmetic inside the circuit. A full opening-and-range circuit with the same public-input count would have the same verifier pairing/MSM operation count. This is a cost proxy, not a measurement of that missing circuit's exact transaction gas: generated code and calldata byte values also affect gas. The stand-in's proving cost is not representative.
\item \textbf{Kernel.} To decompose the cost, a second contract executes only the per-coordinate equations and the reconstruction, without nonce state, enrollment check, challenge, sampling, or threshold verifier.
\item \textbf{Complete in-circuit verifier (\Circ).} A contract implements the same enrollment-and-nonce lifecycle using one Poseidon commitment and a recomputed digest $\chi=\Hash(\ctx\|\id\|\nonce)$, and calls the reference circuit's Groth16 verifier. We measured three authentications at each $\ndim\in\{128,512\}$; replay and proof reuse under a new nonce were rejected.
\item \textbf{Proving.} {Witness, proving, and verification times of the packed reference circuits are medians of five independent cold child processes on \texttt{enine}. Peak memory is \texttt{process.resourceUsage().maxRSS}, converted from KiB to MiB; it is a process maximum and is not baseline-subtracted.}
\end{itemize}
Both homomorphic contracts build $h$ as a known multiple of $g$ so that valid points can be generated deterministically; this is insecure as a commitment key but has the same group-operation cost. {The \Full measurements use a fresh local, single-party threshold setup, whose verified artifacts are reused by \Spot. The \Circ complete-transaction and packed-circuit measurements use a separate fresh local, single-party Groth16 setup. These setups are for benchmarking only.}

{\paragraph*{Security coverage of the benchmarks}
``Complete'' denotes the measured transaction workflow, not a production authentication implementation. In the homomorphic stand-in, choosing $\dist=0$ satisfies the threshold constraints independently of the reconstructed commitment; hence even an exhaustive check can accept a consistent far-vector transcript with that proof. The known relation $h=5g$ independently defeats Pedersen binding. The abstract \Full theorem requires both defects to be absent.

The generated \Circ inputs use public deterministic enrollment/probe salts; its complete-gas driver samples fresh probe salts by reducing 256 random bits modulo the field order. These inputs test constraints and cost, not the uniform-secret-salt privacy hypothesis. Likewise, the contracts use Keccak followed by modular reduction for the challenge or context digest. For BN254's scalar order $r$, $\lfloor2^{256}/r\rfloor=5$: the largest output probability is $6/2^{256}$, and the statistical distance from uniform on $\Fr$ is about $0.03893$. This is not a demonstrated forgery, but it prevents identifying that distribution with the ideal uniform oracle in the exact bounds. A security instantiation would need unbiased field sampling or a proof using its actual distribution. For the measured power-of-two dimensions, a uniform 256-bit word reduced modulo $N$ has no index bias; arbitrary dimensions and real block randomness require separate analysis.}

\subsection{Verification gas at $\ndim{=}128$}

\begin{table*}[t]
\caption{{Measured transaction gas at $\ndim{=}128$ on BN254. The \Full and \Spot rows use a threshold stand-in and do not measure secure complete authentication; the \Circ row uses its reference circuit. Every value is the median of three fresh benchmark transactions. Required rejection controls are defined in the measurement protocol above; ``Tamper'' denotes a modified-response check.}}
\label{tab:gas}
\centering\footnotesize
{
\begin{tabular}{lrrrrll}
\toprule
Strategy & $t$ & Repetitions & Measured tx gas & Calldata (B) & Required controls & Tamper \\
\midrule
\textsc{SpotCheck}, $\eta{=}1$ & 7 & 3 & 4{,}124{,}778 & 127{,}844 & pass & pass \\
\textsc{SpotCheck}, $\eta{=}2$ & 14 & 3 & 5{,}111{,}099 & 127{,}844 & pass & pass \\
\textsc{SpotCheck}, $\eta{=}4$ & 28 & 3 & 7{,}107{,}192 & 127{,}844 & pass & pass \\
\textsc{SpotCheck}, $\eta{=}8$ & 56 & 3 & 11{,}196{,}179 & 127{,}844 & pass & pass \\
\textsc{FullCheck} & 128 & 3 & 22{,}097{,}617 & 127{,}844 & pass & pass \\
\textsc{CircuitCheck} & -- & 3 & 255{,}011 & 356 & pass & n/a \\
\bottomrule
\end{tabular}}
\end{table*}

\begin{table*}[t]
\caption{{\Spot dimension sweep on \texttt{enine}. Each cell is $t$ / median homomorphic stand-in transaction gas from three fresh benchmark transactions, with $t=\min\{N,\lceil\eta\log_2 N\rceil\}$. The four $\eta$ curves contain 28 labelled settings and 22 unique $(N,t)$ pairs: saturation at $t=N$ repeats some pairs at small dimensions.}}
\label{tab:spot-sweep}
\centering\footnotesize
{
\begin{tabular}{rrrrr}
\toprule
$N$ & $\eta{=}1$ & $\eta{=}2$ & $\eta{=}4$ & $\eta{=}8$ \\
\midrule
2   & $1 / 439{,}144$   & $2 / 571{,}255$   & $2 / 571{,}219$   & $2 / 571{,}171$ \\
4   & $2 / 616{,}649$   & $4 / 881{,}173$   & $4 / 881{,}197$   & $4 / 881{,}137$ \\
8   & $3 / 838{,}008$   & $6 / 1{,}240{,}610$ & $8 / 1{,}504{,}201$ & $8 / 1{,}504{,}129$ \\
16  & $4 / 1{,}147{,}981$ & $8 / 1{,}686{,}734$ & $16 / 2{,}762{,}401$ & $16 / 2{,}762{,}329$ \\
32  & $5 / 1{,}637{,}133$ & $10 / 2{,}316{,}909$ & $20 / 3{,}692{,}070$ & $32 / 5{,}328{,}001$ \\
64  & $6 / 2{,}494{,}648$ & $12 / 3{,}320{,}826$ & $24 / 4{,}990{,}872$ & $48 / 8{,}413{,}444$ \\
128 & $7 / 4{,}124{,}778$ & $14 / 5{,}111{,}099$ & $28 / 7{,}107{,}192$ & $56 / 11{,}196{,}179$ \\
\bottomrule
\end{tabular}}
\end{table*}

\begin{table}[t]
\caption{{\Circ complete-transaction and packed-circuit costs on \texttt{enine} (Groth16 over BN254, five public inputs). Timing and max-RSS values are medians from five independent cold processes. Unpacked constraints use reconstructed wrappers with parameters uniquely determined by the generic circuit; the original wrapper files were unavailable.}}
\label{tab:circ}
\centering\footnotesize
{
\begin{tabular}{lrr}
\toprule
 & $N{=}128$ & $N{=}512$ \\
\midrule
Constraints (packed) & 7{,}660 & 27{,}118 \\
Authentication, tx gas & 255{,}011 & 255{,}011 \\
Enrollment, tx gas & 45{,}391 & 45{,}391 \\
Calldata (B) & 356 & 356 \\
Witness, median (ms) & 64.05 & 101.90 \\
Prove, median (ms) & 631.87 & 1{,}031.60 \\
Verify, median (ms) & 316.90 & 314.58 \\
Prove max-RSS, median (MiB) & 822.23 & 1{,}541.20 \\
Constraints (reconstructed unpacked) & 38{,}464 & 151{,}360 \\
Packed constraint reduction & 80.1\% & 82.1\% \\
\bottomrule
\end{tabular}}
\end{table}

\begin{figure*}[t]
\centering
\includegraphics[width=\textwidth]{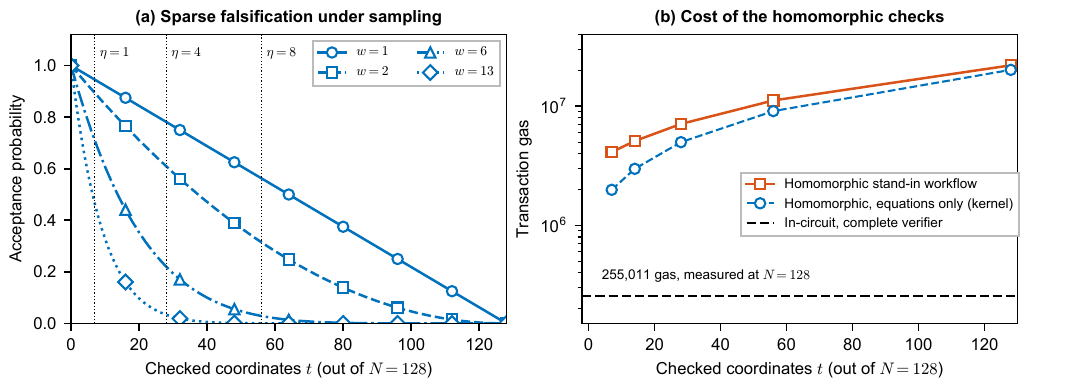}
\caption{{Sampling and cost at $\ndim{=}128$. (a) Probability that a transcript falsified in $\nbad$ coordinates passes a sample of $\nsamp$ coordinates (Proposition~\ref{prop:accept}). (b) Transaction gas versus examined coordinates. The homomorphic stand-in workflow curve combines the four \Spot settings and the \Full measurement; the \Circ line shows its complete-transaction cost. The equations-only kernel series uses historical measurements.}}
\label{fig:cost}
\end{figure*}

{Tables~\ref{tab:gas}--\ref{tab:circ} report measured transaction costs across the tested dimensions; the homomorphic values describe the stand-in workflow. Figure~\ref{fig:cost} shows the acceptance probability and transaction gas at $\ndim=128$ as the number of examined coordinates changes.}
\begin{itemize}
\item At $N=128$, the \Full stand-in workflow costs $22{,}097{,}617$ gas. Checking seven coordinates with \Spot ($\sfac=1$) reduces this to $4{,}124{,}778$ gas, a saving of $81.3\%$. This is a substantial workflow saving, but Proposition~\ref{prop:accept} gives a $94.5\%$ miss probability for one falsified coordinate under uniform sampling. The sampled workflow also remains $16.2\times$ the complete \Circ cost of $255{,}011$ gas; the exhaustive workflow is $86.7\times$ that cost.
\item {Using the recorded equations-only kernel values solely as a decomposition reference, the measured stand-in workflow adds $1.86$ to $2.14$ million gas. The transcript is $127{,}844$ bytes at every sampling factor because the prover cannot know $\Samp$ in advance; sampling saves equation evaluations but not data. A commitment to the first messages with per-coordinate openings could shrink the calldata of \Spot; we did not evaluate it.}
\item {Across the $N=128$ homomorphic stand-in transaction endpoints, each additional examined coordinate costs approximately $148{,}536$ gas, comparable in scale to the $109{,}800$ gas assigned by EIP-1108~\cite{eip1108} to its $18$ multiplications and $12$ additions. The historical kernel decomposition attributes about $1.0$ million gas at $\ndim{=}128$ to reconstruction.}
\item {Complete \Circ transactions cost $255{,}011$ gas at both measured dimensions, $128$ and $512$. Enrollment costs $45{,}391$ gas for one commitment, compared with $192{,}709$ gas for the digest of $\ndim$ Pedersen commitments.}
\end{itemize}

{\color{black}\paragraph*{Group size and cost}
Gas figures for the homomorphic checks depend on the group: a reduced group (for instance a $26$-bit modulus, Table~\ref{tab:params}) makes the arithmetic cheaper but removes binding, so such figures do not describe a secure configuration. We therefore use the earlier equations-only kernel measurements only to decompose cost, not to estimate secure homomorphic verification. A single-proof design that verifies Pedersen commitments \emph{inside} the circuit pays for non-native group arithmetic; \Circ avoids this by committing with a SNARK-native hash.}

\subsection{Prover cost and setup}
{The packed reference circuit has median proving times of $631.87$\,ms ($\ndim{=}128$) and $1{,}031.60$\,ms ($\ndim{=}512$), with median process max-RSS of $822.23$ and $1{,}541.20$\,MiB. Median witness times are $64.05$ and $101.90$\,ms, and median verification times are $316.90$ and $314.58$\,ms. All ten proofs verified successfully. Relative to the reconstructed unpacked wrappers ($38{,}464$ and $151{,}360$ constraints), packing reduces constraints by $80.1\%$ and $82.1\%$. We exclude the homomorphic threshold stand-in's proving times because it omits the Pedersen opening and uses an insecure commitment key; those timings cannot characterize secure \Full or \Spot proving. All three strategies require circuit-specific Groth16 setup.}

\subsection{Local proof-processing overhead}\label{sec:cost:local}
\begin{figure*}[t]
\centering
\includegraphics[width=\textwidth]{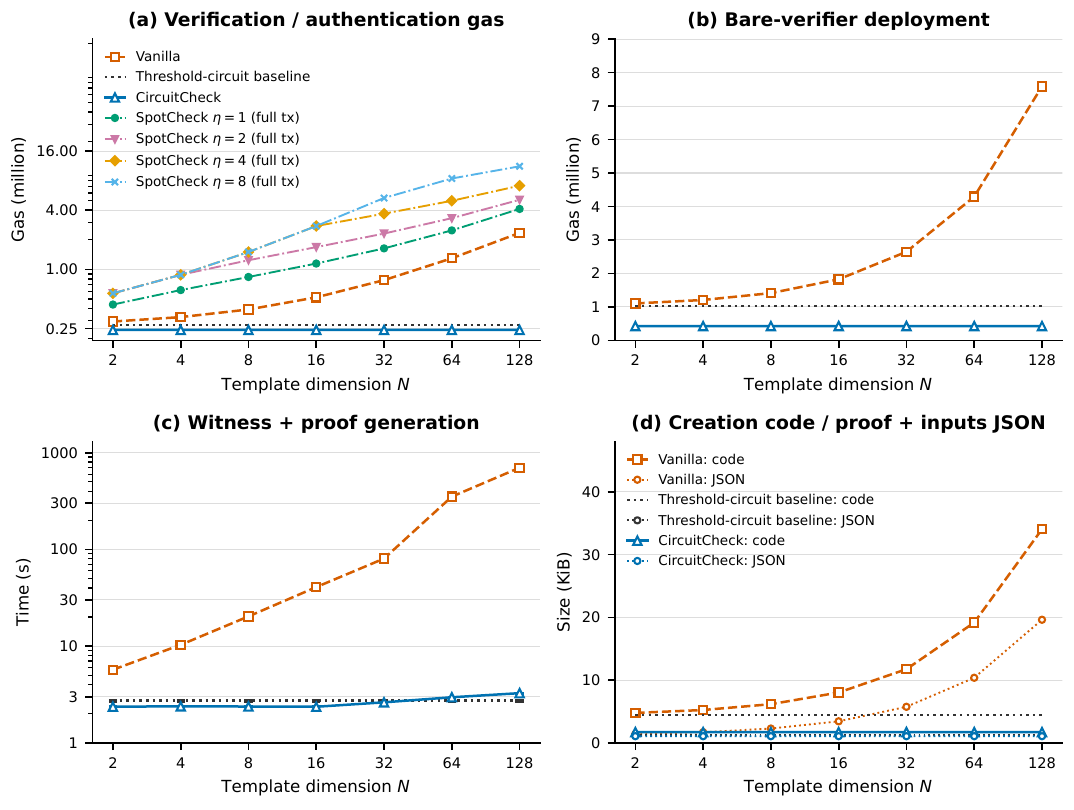}
\caption{{Measured costs on \texttt{enine} for $N\in\{2,4,8,16,32,64,128\}$. (a) Bare Groth16 verification gas for Vanilla, the threshold-circuit baseline, and CircuitCheck, alongside \Spot stand-in transaction gas for all 28 $(N,\eta)$ settings ($\eta=1,2,4,8$, three-transaction medians). The \Spot costs use the benchmark stand-in described above. Panels~(b)--(d) compare the two Groth16 component baselines with \Circ. (c) Witness-plus-proof time, shown as mean $\pm$ sample SD over five samples for Vanilla and the threshold-circuit baseline and ten for CircuitCheck. Toolchains, startup accounting, and thread policies differ as detailed in the text. (d) Creation bytecode and serialized proof-plus-public-input JSON sizes. Vanilla at $N=128$ requires a relaxed contract-size limit (34,837-byte runtime); all CircuitCheck verifiers satisfy EIP-170. The threshold-circuit baseline repeats one dimension-independent threshold/opening circuit.}}
\label{fig:isc}
\end{figure*}

Figure~\ref{fig:isc} compares local proof-processing costs using two Groth16 component baselines: Vanilla, whose circuit depends on the vector dimension, and the threshold-circuit baseline, which reuses one dimension-independent threshold/opening circuit. The latter isolates the threshold-proof component of the homomorphic verification path. Tables~\ref{tab:gas} and~\ref{tab:circ} report the homomorphic stand-in transaction costs and complete \Circ reference transactions, respectively. These component timings do not measure complete secure \Full or \Spot proving.

{The figure also includes \Circ across seven dimensions and the \Spot gas sweep. Vanilla and the threshold-circuit baseline reuse their compiled circuits, proving and verification keys, and matching Solidity verifiers; \Circ uses fresh local benchmark keys. Each Vanilla circuit received a minimal repair removing a shadowing \texttt{res} declaration (Appendix~\ref{app:prov}). The \Circ series uses ten cold witness-plus-proof samples and five bare-verifier transactions per dimension, following one warm-up for each measurement series.}

{Witness-plus-proof times for Vanilla and the threshold-circuit baseline are five-run means with sample SDs: the threshold-circuit baseline takes $2.728\pm0.070$\,s, while Vanilla rises from $5.701\pm0.262$\,s at $N=2$ to $699.007\pm13.314$\,s at $N=128$. The measurement protocol uses default Rayon threading through $N=32$ and one Rayon thread for $N\in\{64,128\}$ and the threshold-circuit baseline. Gas is $271{,}200\pm0$ for the threshold-circuit baseline ($n=30$ transactions), $295{,}244\pm0$ for Vanilla at $N=2$, and $2{,}354{,}712\pm0$ at $N=128$ (each $n=5$ transactions). These gas repetitions reuse a cached proof.}

{CircuitCheck's mean witness-plus-proof time ranges from $2.367$\,s at $N=2$ to $3.263$\,s at $N=128$. Its mean bare-verifier gas ranges from $242{,}256$ to $242{,}280$, and its creation bytecode is $1{,}751$--$1{,}752$ bytes. This grid uses Circom~2.2.3/snarkjs~0.7.6 on CPU~0 with single-thread environment settings; Vanilla and the threshold-circuit baseline use ZoKrates~0.8.8 with the Rayon settings above. CircuitCheck timing starts inside a cold Node child process and includes witness calculation and proof/public-input JSON writes. For Vanilla and the threshold-circuit baseline, timing wraps the witness and proof commands and includes process startup. These differences preclude attributing the observed timing gap solely to the algorithms. Serialized sizes preserve each tool's formatting: one hexadecimal proof-with-inputs file for ZoKrates, and the combined decimal proof and public-input files for snarkjs. They do not measure a common binary encoding.}

{The means for Vanilla, the threshold-circuit baseline, and CircuitCheck contain only newly collected samples; the \Spot overlay retains the original medians for all 28 measured configurations. Deployment controls confirm that Vanilla's 34,837-byte runtime at $N=128$ exceeds the 24,576-byte EIP-170 limit; deployment and verification succeed when the local chain permits unlimited contract size. Detailed records are provided in Appendix~\ref{app:prov}.}

\subsection{Quantization and recognition quality}\label{sec:quant}

The shared biometric front end supplies quantized integer vectors to all three strategies. Before commitment, each embedding is $\ell_2$-normalized, mapped to the integer range for $b$ bits, and rounded; Euclidean distance is then computed from the quantized coordinates. We evaluate this front end on two datasets, and Table~\ref{tab:quant} reports the results. FEI contains $200$ identities with $14$ images per identity, acquired under controlled conditions~\cite{thomaz2010fei}. Labeled Faces in the Wild (LFW) contains faces acquired under unconstrained conditions~\cite{huang2007lfw}; we use its Funneled alignment~\cite{huang2007funneling}.

Both datasets use the same quantizer, deterministic identity-disjoint split algorithm, and evaluation policy. The validation/test splits contain $40/160$ identities for FEI and $1{,}149/4{,}600$ for LFW. Before face detection, each partition receives a fixed ordered list of $3{,}000$ genuine and $3{,}000$ impostor image pairs (P6000). Detection failures remove pairs without replacement or resampling, and each model's eligible pairs are reused at every bit width. LFW sampling is globally uniform over unordered image pairs, yielding image-pair-weighted estimates. For each model and dataset, the threshold maximizes balanced accuracy on eligible float validation pairs and remains fixed for float and all quantized test settings. This common evaluation protocol differs from the official LFW 10-fold protocol.

We evaluate Qualcomm AI Hub MobileFaceNet-128D~\cite{qualcommMobileFaceNet} at float precision and $4$, $6$, $8$, and $12$ bits, and \texttt{facenet-pytorch} InceptionResnetV1 pretrained on VGGFace2~\cite{facenetPytorch} at float precision and $4$, $6$, $7$, and $9$ bits. On FEI, MobileFaceNet has $5{,}935$ eligible test pairs and InceptionResnetV1 has $5{,}921$; both have $5{,}857$ eligible validation pairs. On LFW, the corresponding test/validation counts are $5{,}949/5{,}980$ for MobileFaceNet and $6{,}000/6{,}000$ for InceptionResnetV1. Accuracy, recall, false-accept rate (FAR), false-reject rate (FRR), and AUC are evaluated on eligible test pairs. FAR denotes the fraction of accepted impostor pairs; FRR equals one minus recall.

\begin{table*}[t]
\caption{Recognition quality under quantization on controlled-acquisition FEI and unconstrained-acquisition LFW Funneled (percent, except AUC). $\Delta$ Acc. is the within-model, within-dataset accuracy change from float in percentage points. ``Pairs'' is the model-specific eligible test count after face detection. Each threshold is selected from float validation distances only and held fixed across all listed bit widths.}\label{tab:quant}
\centering\scriptsize\setlength{\tabcolsep}{2.3pt}
\begin{tabular}{@{}llrrrrrrr@{}}
\toprule
Model & Setting & Pairs & Acc. & $\Delta$ Acc. & Rec. & FAR & FRR & AUC \\
\midrule
\multicolumn{9}{l}{FEI Face Database --- controlled acquisition} \\
MobileFaceNet (128) & float & 5{,}935 & 97.88 & -- & 97.24 & 1.48 & 2.76 & 0.99654 \\
MobileFaceNet (128) & 4-bit & 5{,}935 & 90.51 & -7.36 & 81.09 & 0.07 & 18.91 & 0.99282 \\
MobileFaceNet (128) & 6-bit & 5{,}935 & 97.78 & -0.10 & 96.97 & 1.42 & 3.03 & 0.99641 \\
MobileFaceNet (128) & 8-bit & 5{,}935 & 97.96 & +0.08 & 97.37 & 1.45 & 2.63 & 0.99656 \\
MobileFaceNet (128) & 12-bit & 5{,}935 & 97.88 & +0.00 & 97.27 & 1.52 & 2.73 & 0.99654 \\
InceptionResnetV1 (512) & float & 5{,}921 & 97.67 & -- & 95.50 & 0.17 & 4.50 & 0.98531 \\
InceptionResnetV1 (512) & 4-bit & 5{,}921 & 86.30 & -11.37 & 72.65 & 0.07 & 27.35 & 0.98007 \\
InceptionResnetV1 (512) & 6-bit & 5{,}921 & 97.21 & -0.46 & 94.49 & 0.07 & 5.51 & 0.98532 \\
InceptionResnetV1 (512) & 7-bit & 5{,}921 & 97.62 & -0.05 & 95.33 & 0.10 & 4.67 & 0.98530 \\
InceptionResnetV1 (512) & 9-bit & 5{,}921 & 97.69 & +0.02 & 95.50 & 0.13 & 4.50 & 0.98529 \\
\midrule
\multicolumn{9}{l}{LFW Funneled --- unconstrained acquisition} \\
MobileFaceNet (128) & float & 5{,}949 & 98.67 & -- & 97.34 & 0.00 & 2.66 & 0.98495 \\
MobileFaceNet (128) & 4-bit & 5{,}949 & 83.73 & -14.94 & 67.46 & 0.00 & 32.54 & 0.98618 \\
MobileFaceNet (128) & 6-bit & 5{,}949 & 98.62 & -0.05 & 97.24 & 0.00 & 2.76 & 0.98505 \\
MobileFaceNet (128) & 8-bit & 5{,}949 & 98.67 & +0.00 & 97.34 & 0.00 & 2.66 & 0.98493 \\
MobileFaceNet (128) & 12-bit & 5{,}949 & 98.67 & +0.00 & 97.34 & 0.00 & 2.66 & 0.98495 \\
InceptionResnetV1 (512) & float & 6{,}000 & 96.72 & -- & 94.40 & 0.97 & 5.60 & 0.98050 \\
InceptionResnetV1 (512) & 4-bit & 6{,}000 & 92.72 & -4.00 & 85.60 & 0.17 & 14.40 & 0.97559 \\
InceptionResnetV1 (512) & 6-bit & 6{,}000 & 96.87 & +0.15 & 94.30 & 0.57 & 5.70 & 0.98017 \\
InceptionResnetV1 (512) & 7-bit & 6{,}000 & 96.75 & +0.03 & 94.37 & 0.87 & 5.63 & 0.98046 \\
InceptionResnetV1 (512) & 9-bit & 6{,}000 & 96.72 & +0.00 & 94.40 & 0.97 & 5.60 & 0.98046 \\
\bottomrule
\end{tabular}

\end{table*}

Table~\ref{tab:quant} shows the strongest LFW degradation at $4$ bits: accuracy falls from $96.7167\%$ to $92.7167\%$ for InceptionResnetV1 and from $98.6720\%$ to $83.7284\%$ for MobileFaceNet. With $2{,}000$ identity-cluster bootstrap replicates, the paired $4$-bit accuracy-change intervals exclude zero: $[-5.7490,-1.3821]$ percentage points for InceptionResnetV1 and $[-21.4059,-6.0542]$ for MobileFaceNet. Every evaluated setting at $6$ bits or above is within $0.15$ percentage points of its LFW float baseline; Table~\ref{tab:quant} gives the per-width results.

FEI shows the same qualitative pattern: $4$-bit quantization reduces accuracy by $7.36$ percentage points for MobileFaceNet and $11.37$ for InceptionResnetV1, whereas every evaluated setting at $6$ bits or above is within $0.46$ points of float. The lower FAR at $4$ bits accompanies a higher rejection rate for genuine pairs because the threshold remains fixed as distances shift. These comparisons assess the biometric front end within each model and dataset; they do not establish verification-strategy soundness. All evaluated widths fit the $16$-bit coordinate range of \Circ.

FEI has a homogeneous white background and a limited subject population. LFW offers more variable acquisition conditions, but our evaluation uses a single deterministic split and image-pair-weighted sampling. The pipelines also differ in detection and preprocessing, limiting comparisons across models. The observed quantization effects therefore cannot be extrapolated to other populations, sensors, presentation attacks, or deployment-scale low-FAR operating points.

\subsection{Threats to validity}
All systems-cost and recognition experiments use the single host specified in \S\ref{sec:cost}; recognition inference runs on the CPU. Gas and proving times come from separate measurement programs and are reported separately. The homomorphic contracts are benchmark implementations without gas optimization, and their threshold stand-in supports only the verifier-cost comparison qualified above. The Poseidon commitment of \Circ requires collision resistance of its fixed-format hash compression and a separate message-hiding assumption (\S\ref{sec:prelim}). The implementations and security arguments have not undergone an independent audit.

\section{Related Work}\label{sec:related}

\paragraph*{In-circuit verification}
Kothari et al.~\cite{kothari2026} implement in-circuit fingerprint verification using a cancelable enrollment template committed with Poseidon and stored encrypted off chain. Their Groth16 circuit recomputes a division-free matching rule, range-checks every private value, and binds a session nonce as a public signal. They report $255{,}275$ gas for verification, a proof of three group elements (about $800$ bytes as reported), and recognition results on the twelve FVC2002--2006 databases under a held-out protocol. Our \Circ figure of $255{,}011$ gas covers a complete transaction at $\ndim{=}128$ with five public inputs; their figure uses a different circuit and measurement procedure, so the two numbers are not a controlled comparison. Their work also evaluates cancelable template protection and encrypted storage. They state that the scheme must be restricted to one-to-one verification and that the query enters the circuit unconstrained.

\emph{Identity and template protection.} A Web 3.0 survey identifies identity management as a core service workflow and places blockchain among its enabling technologies~\cite{cao2026web3}. Prior systems protect stored templates or offer private matching~\cite{salem2024blockchain,sarier2022privacy,zeng2025ppbiometric}, and blockchain identity schemes improve auditability and cross-domain verification~\cite{gao2018blockchain,gilani2020survey,yang2020zkdid}. Zero-knowledge biometric schemes reduce plaintext exposure~\cite{mao2025zkp}. Their use of a circuit leaves a separate design question: which commitment and matching operations should it prove? Our comparison examines homomorphic checks outside the circuit and a complete relation using SNARK-native commitments; Section~\ref{sec:cost:local} reports component timings with different toolchains and timing boundaries.

\emph{Proofs of matching.} zkSABER~\cite{chinen2025zksaber} combines cosine matching, anonymous set membership, and a nullifier in one circuit, with a public statement comprising a root, a nullifier, and a scaled threshold. BAZKP~\cite{mao2023bazkp} combines a Euclidean-distance zero-knowledge proof with a range proof to show that a committed fingerprint vector lies within a distance threshold. It uses a user-and-cloud architecture rather than a contract. ZK-SERIES~\cite{reijsbergen2025zkseries} authenticates time-series biometrics based on shaking and blowing. It uses zero-knowledge multiplication proofs and batched range proofs on smartphones to establish that recent readings lie within a distance threshold of baseline readings; its range proofs address the gap noted in Remark~\ref{rem:range}. BioZKFHE~\cite{xin2026biozkfhe} addresses outsourced one-to-many encrypted biometric identification: BGV evaluates similarities, a threshold committee checks opened proof instances, and the contract enforces the committee's attestations. Its verification target is encrypted similarity computation; our analysis compares coordinate sampling with a complete circuit proof for a one-to-one committed match. Our comparison asks whether the cost of exhaustive homomorphic verification can be reduced by sampling its coordinate checks without losing the modular matching guarantee. We analyze \Spot as a variant for answering that question and compare the resulting security and cost constraints with a complete circuit proof of integer matching.

\section{Discussion}\label{sec:discussion}

\subsection{Verification strategy selection}
Table~\ref{tab:choice} summarizes what each strategy verifies and the measured cost at $\ndim=128$. The homomorphic cost columns refer to the stand-in workflow, while the security rows state properties of the abstract protocols. All three designs use Groth16 under the setup and proof assumptions in \S\ref{sec:model}; the assumptions row lists additional strategy-specific requirements.

\begin{table*}[t]
\caption{Strategy comparison at $\ndim{=}128$ on BN254. Gas values are measured transactions from Tables~\ref{tab:gas} and~\ref{tab:circ}: the \Full and \Spot columns use a threshold stand-in, whereas \Circ uses its reference circuit.}\label{tab:choice}
\centering\small
\begin{tabular}{@{}p{2.4cm}p{3.2cm}p{3.0cm}p{3.0cm}p{4.6cm}@{}}
\toprule
 & \Full & \Spot ($\sfac{=}1$) & \Spot ($\sfac{=}8$) & \Circ \\
\midrule
Sparse falsification & checks every coordinate & passes with $94.5\%$ for $\nbad{=}1$ & passes with $56.3\%$ for $\nbad{=}1$ & proves the complete relation \\
Coordinate range enforced & no & no & no & yes ($16$ bits) \\
Measured tx gas & $22.1$\,M & $4.12$\,M & $11.2$\,M & $0.255$\,M at both measured dimensions \\
Calldata & $128$\,kB & $128$\,kB & $128$\,kB & $356$\,B \\
Enrollment record (model) & $\ndim$ commitments & same & same & one commitment \\
Measured enrollment & digest of $\ndim$ commitments; $193$\,k gas & same & same & one commitment; $45$\,k gas \\
Prover work & $O(\ndim)$ exponentiations $+$ threshold proof & same & same & one circuit proof ($0.632$\,s, $822.23$\,MiB) \\
Setup & Groth16 (threshold) & Groth16 (threshold) & Groth16 (threshold) & Groth16 (relation) \\
Additional assumptions & Pedersen binding, ROM, full-state extraction & Pedersen binding, ROM, uniform sampling & same & Poseidon binding and hiding; context-hash collision resistance \\
\bottomrule
\end{tabular}
\end{table*}

\Full establishes the modular matching relation under Proposition~\ref{prop:full} and rejects every transcript containing a failed coordinate equation. It is the exhaustive baseline, with coordinate-check and reconstruction work linear in $\ndim$. Its measured stand-in workflow costs $22.1$ million gas at $\ndim=128$; the cost of a secure complete implementation remains unmeasured.

\Spot tests whether sampling can reduce this cost while retaining the homomorphic structure. At $\ndim=128$, checking seven coordinates saves $81.3\%$ of the stand-in gas, but a single falsified coordinate passes with probability $94.5\%$ under uniform sampling. Limiting this attack's acceptance probability to $\varepsilon$ requires checking at least $(1-\varepsilon)\ndim$ coordinates, leaving little scope for savings in coordinate checks when $\varepsilon$ is small. The evaluated sampled workflows also retain the full $128$\,kB transcript and cost $16$ to $44$ times the \Circ reference transaction. The value of this comparison is to identify the limit of coordinate sampling as a replacement for exhaustive verification in the specified protocol.

For a blockchain-backed identity service, the transaction is the public evidence available to a relying application. The full transcript required by \Spot is therefore a data-transfer cost even when the contract verifies only a few coordinates. The \Circ transaction, with $356$ bytes of calldata, gives a smaller on-chain interface for the evaluated dimensions, while moving work to the prover. Network-wide throughput also depends on blockchain architecture and consensus design, including sharding~\cite{cao2023methodology,cao2026shards}; we do not measure those effects.

\paragraph*{Recommended instantiation}
These results motivate changing the verification structure rather than relying on sublinear coordinate sampling. For the evaluated integer matching task, we recommend \Circ: a salted, role-separated Poseidon commitment to packed $16$-bit coordinates, a Groth16 circuit that checks both vectors and the distance threshold, and a digest $\chi$ binding identity, nonce, and deployment context. Packing reduces constraints by $80.1\%$ at $\ndim=128$ and $82.1\%$ at $\ndim=512$ relative to the reconstructed unpacked wrappers. The complete transaction costs approximately $255$ thousand gas in both measured dimensions, with $356$ bytes of calldata. On the evaluated workstation, median proving time is $0.632$\,s at $\ndim=128$ and $1.032$\,s at $\ndim=512$, with process max-RSS of $822.23$ and $1{,}541.20$\,MiB. These measurements characterize the client resources required by the recommendation; secure homomorphic proving costs remain unmeasured. The design also requires the Poseidon assumptions stated in Section~\ref{sec:security} and a dimension-specific setup. \Full remains an alternative for the modular matching relation when homomorphic commitments are required and verification cost is less restrictive, subject to the parameter and extraction assumptions of Table~\ref{tab:params} and Proposition~\ref{prop:full}. Enforcing the bounded-integer relation of \Circ would additionally require coordinate-range proofs.

\subsection{Relation to biometric capture}
The protocols verify a relation between committed vectors. Binding the probe to a live sensor reading and enforcing unique enrollment require additional mechanisms. A party holding the enrollment secrets can submit the enrolled template as its probe, so deployment also requires secure secret storage and an authenticated capture path. Kothari et al.~\cite{kothari2026} identify the same issue: an unconstrained query allows authentication with a stolen key without presenting the finger.

\subsection{Limitations}
The security results depend on the extraction and commitment assumptions in Section~\ref{sec:security}. The privacy analysis covers the protocol messages for one honest enrollment and successful authentication. Registration authorization, front-running, and leakage across repeated decisions require separate analysis. The reference registry also requires nonzero commitments, canonical policy inputs, and a verifier matching the configured dimension. Sampling guarantees additionally require the randomness model of Proposition~\ref{prop:attempts}; a deployment using block-derived randomness must address proposer bias and transaction ordering, for example with a suitable VRF or commit--reveal scheme.

The homomorphic benchmarks use a known discrete-log relation between generators and a threshold circuit that omits the Pedersen-opening check. Their gas measurements therefore characterize the implemented verification workflow; secure homomorphic proving costs remain unmeasured. All benchmark setups are single-party and unsuitable for deployment. Figure~\ref{fig:isc} compares implementations with different toolchains, timing boundaries, and threading policies, as specified in Section~\ref{sec:cost:local}. Vanilla at $N=128$ also exceeds EIP-170 and requires the local contract-size limit to be disabled.

The recognition results use one identity-disjoint split per dataset. Detection failures introduce model-specific selection effects, and LFW pair sampling weights image pairs rather than identities equally. The bootstrap intervals quantify uncertainty within the selected split. Thresholds are selected on float validation data and then fixed, so the results do not establish performance at deployment-scale false-acceptance rates or under domain shift. Different detection and preprocessing pipelines also prevent a controlled comparison of the two model architectures. Presentation attacks, revocation, and demographic variation remain open evaluation questions.

\section{Conclusion}\label{sec:conclusion}

\sysname{} examines how to reduce the on-chain cost of publicly verifiable committed biometric matching. Under the stated assumptions, \Full provides knowledge soundness for the modular matching relation, but its coordinate checks and reconstruction require work linear in dimension; integer matching additionally requires coordinate-range proofs. \Spot tests a natural cost-reduction attempt within this homomorphic structure. At $\ndim=128$, checking seven coordinates reduces stand-in gas from $22.1$ million to $4.1$ million, an $81.3\%$ saving, but leaves the full $127{,}844$-byte calldata. The single-coordinate attack passes with probability $1-\nsamp/\ndim$ under uniform sampling, so negligible error requires nearly exhaustive checking. This limits the saving available from sampling while retaining the exhaustive guarantee. \Circ changes the commitment and proof structure to establish the complete integer relation, with approximately $255$ thousand gas and $356$ bytes of calldata for its complete reference transaction. Together, the security analysis and measurements favor this design for the evaluated task, subject to its commitment and setup assumptions and client proving requirements.

Across the evaluated FEI and LFW pipelines, every tested width of $6$ bits or more preserves accuracy within $0.46$ percentage points of the corresponding float baseline and fits the circuit's $16$-bit coordinate range. Four-bit quantization reduces accuracy and recall at the fixed thresholds; both paired LFW accuracy-change intervals exclude zero. These results support moderate-width integer embeddings for the shared biometric front end, independently of the verification strategy. Deployment must additionally connect the committed probe to an authenticated live capture, as discussed in Section~\ref{sec:discussion}.

\section*{Data and Code Availability}
Code and compact experimental results are available in the \href{https://github.com/xiaolin9595/BIOZERO-Verifiable-On-Chain-Biometric-Authentication-}{\sysname{} repository}. Release \texttt{v1.0.0}, at commit \texttt{53385d0c600c}, is archived at \href{https://doi.org/10.5281/zenodo.23758484}{doi:10.5281/zenodo.23758484}. The package includes contracts, circuits, pinned toolchains, analysis scripts, aggregate measurements, and provenance records. Appendix~\ref{app:prov} identifies the sources and measurement protocols. The CircuitCheck grid summary and plotting updates accompanying this revision postdate the archived release.

The public package excludes biometric images and templates, embeddings, per-pair records, model weights, and generated proving keys. FEI data remain subject to the provider's research-use and redistribution terms. Experimental keys are for benchmarking only.

\appendices
\section{Proofs and Derivations}\label{app:proofs}

The derivations assume the protocol model and cryptographic assumptions of \S\ref{sec:model}.

\paragraph*{Completeness of \eqref{eq:v1}--\eqref{eq:v2}}
For an honest transcript, $g^{z^{\mathrm{m}}}h^{z^{\mathrm{r}}}=g^{b_1+\chal m}h^{b_2+\chal\blind}=\alpha\,(g^{m}h^{\blind})^{\chal}=\alpha\,(c^{\anc})^{\chal}$. For \eqref{eq:v2}, with $c^{\bas}=g^{m'}h^{\blind'}$ and $c^{\lrel}=g^{m\,m'}h^{\blind^{\lrel}}$ (so that $m=m^{\anc}$ and $m'=m^{\bas}$, $\blind'=\blind^{\bas}$):
$(c^{\bas})^{b_1+\chal m}\,h^{b_3+\chal(\blind^{\lrel}-\blind' m)}
=\bigl((c^{\bas})^{b_1}h^{b_3}\bigr)\,\bigl(g^{m m'}h^{\blind' m+\blind^{\lrel}-\blind' m}\bigr)^{\chal}=\beta\,(c^{\lrel})^{\chal}$.

\paragraph*{Proof of Proposition~\ref{prop:accept}}
If $\nsamp>\ndim-\nbad$, every sample intersects $W$ and the miss probability is zero. Otherwise, counting subsets gives the ratio in \eqref{eq:sampling-pass}. For $\nsamp>0$, that ratio factors as follows:
\begin{equation}\label{eq:sampling-product}
\Pass(\ndim,\nsamp,\nbad)
=\prod_{j=0}^{\nsamp-1}\frac{\ndim-\nbad-j}{\ndim-j}.
\end{equation}
Every factor is nonnegative and at most $(\ndim-\nbad)/\ndim$, proving the bound. For $\nsamp=0$, the empty sample misses with probability one. For $\nbad=1$, the ratio is $(\ndim-\nsamp)/\ndim$, including the full-sampling endpoint. Requiring it to be at most $\varepsilon$ gives $\nsamp\ge(1-\varepsilon)\ndim$. The capped logarithmic schedule has $\nsamp/\ndim\to0$ for fixed $\sfac$.

\paragraph*{Proof of Proposition~\ref{prop:single}}
Let $\dist=\dist(\vm{e},\vm{p})$ and $\blind_{\dist}$ be the honest values, and fix $\dist^{\ast}\in[0,\thr]$. Since $\ord$ is an odd prime, $2$ is invertible modulo $\ord$. Set $\Delta=(\dist-\dist^{\ast})\cdot2^{-1}\bmod\ord$ and choose $\blind'\in\Zq$. Replace $\cc{ep}{j}$ by $\tilde c_j=\cc{ep}{j}\cdot\Com_{\ck}(\Delta;\blind')$ and keep all other commitments. By \eqref{eq:recon},
\begin{equation}\label{eq:single-coordinate-shift}
\begin{aligned}
\Recon(\dots\tilde c_j\dots)&=\Com_{\ck}(\dist;\blind_{\dist})\cdot\Com_{\ck}(\Delta;\blind')^{-2}\\
&=\Com_{\ck}(\dist^{\ast};\blind_{\dist}-2\blind').
\end{aligned}
\end{equation}
The prover knows this opening and produces an honest Groth16 proof of the threshold statement for $\dist^{\ast}\in[0,\thr]$. The challenge \eqref{eq:chal} is computed after all commitments are fixed, so nothing is circular. For $i\ne j$ the Sigma arguments are honest. At $i=j$, after computing the challenge, the prover deliberately makes the first $\mathrm{ep}$ equation fail: choose responses arbitrarily and, if that equation holds, increment $z^{\mathrm{ep},\mathrm{r}}_j$ by one. Since $h\ne1$, it then fails, including when $\chal=0$. Responses are not part of the challenge input. Exactly coordinate $j$ therefore fails the per-coordinate checks, so acceptance is equivalent to $j\notin\Samp$. A toy-group check of the distance-shift identity above is in \texttt{analysis/verify\_claims.py}.

\paragraph*{Proof of Proposition~\ref{prop:attempts}}
{Write $P=\Pass(\ndim,\nsamp,\nbad)$. Conditional on any prior failure history, a fresh uniform subset misses the current already fixed size-$\nbad$ set with probability $P$. By the chain rule, all attempts fail with probability $(1-P)^{\nattempt}$; adaptive choice of the set does not require unconditional independence of the transcripts.

For predictable sampling, first fix an honest matching commitment tuple and first messages, compute the challenge and sample, and then corrupt responses only outside that sample. Responses do not enter the challenge, so the sample is unchanged. Changing a product commitment to shift the distance changes the challenge and can change the sample. For that attack, fix the shifted product commitment and sample fresh first-message masks, recomputing the challenge, sample, and responses until the bad coordinate is missed. Under fresh independent ideal sample outputs, the success probability is $1-(1-P)^K$ after $K$ candidates ($\nbad=1$). Neither construction yields a miss at $\nsamp=\ndim$.}

\paragraph*{Proof of Proposition~\ref{prop:full}}
{\emph{Soundness under the stated extraction assumption.} Put $Q=q_{\Hash}+1$. Wrap $\Adv$ in an algorithm that runs it and then verifies its output. The wrapper numbers distinct random-oracle queries, caches repeated queries, and makes the final Fiat--Shamir query itself if $\Adv$ omitted it. On acceptance it returns the index of that query; otherwise it returns zero. Its acceptance probability is exactly $p$, and it uses at most $Q$ queries, including the verifier's query. Unused random-answer slots can be padded. Thus guessing a challenge without querying the oracle is included, rather than declared impossible.

The general forking lemma~\cite[Lemma~1]{bellare2006forking} gives a fork probability at least $p(p/Q-1/|\Chals|)$. Replay uses the same random tape and answer prefix, resampling the indexed answer and subsequent answers. In a successful fork both runs accept and return the same index with different challenges. Their indexed query strings therefore agree. With canonical, unambiguous encoding, that string fixes every commitment and first message in \eqref{eq:chal}; one fork supplies two transcripts for all $3\ndim$ relations. This argument also covers a verifier-added final query.

To combine extraction, define a Groth16 adversary $\mathcal B$ that internally runs $\Adv$, lazy-samples the oracle, performs wrapper verification, and outputs the first branch's threshold statement and proof, retaining its transcript and query index as side state. Its coins include the prover tape and oracle-answer vector. Apply full-state knowledge soundness in the sense of Groth~\cite[Sec.~2.2]{groth2016size} to this composed adversary, with the benign auxiliary inputs and independently generated CRS required in \S\ref{sec:model}. Its extractor produces a witness for that same output, with accepting-but-invalid-witness probability at most $\delta_{\mathsf{G16}}$.

Save the first branch's input, coins, and oracle table. Execute $\mathcal B$ and its state-access extractor on an isolated copy using those same coins; the copied run reproduces the identical $(c'_{\dist},\thr,\pi)$ and leaves the saved replay state unchanged. Then perform the fork above on the original saved state. Thus threshold extraction and forking concern the same first transcript, under the full-state extraction assumption for the composed adversary.}

Fix such a pair and write $\Delta\chal=\chal-\chal'\ne0$ in the field $\Zq$, and $\Delta z$ for the difference of corresponding responses. For each coordinate $i$ and label $\lrel$, dividing the two transcript identities in \eqref{eq:v1} gives $g^{\Delta z^{\lrel,\mathrm{m}}_i}h^{\Delta z^{\lrel,\mathrm{r}}_i}=(\cc{\anc(\lrel)}{i})^{\Delta\chal}$; raising both sides to $\Delta\chal^{-1}$ yields the opening
\begin{equation}\label{eq:fork-anchor-opening}
\cc{\anc(\lrel)}{i}=g^{m^{\lrel}_i}h^{\rho^{\lrel}_i},\qquad m^{\lrel}_i=\Delta z^{\lrel,\mathrm{m}}_i/\Delta\chal,\ \ \rho^{\lrel}_i=\Delta z^{\lrel,\mathrm{r}}_i/\Delta\chal .
\end{equation}
Dividing the two identities in \eqref{eq:v2} and raising both sides to $\Delta\chal^{-1}$ gives $(\cc{\bas(\lrel)}{i})^{m^{\lrel}_i}h^{u^{\lrel}_i}=\cc{\lrel}{i}$, where $u^{\lrel}_i=\Delta z^{\lrel,\mathrm{u}}_i/\Delta\chal$. The same $m^{\lrel}_i$ occurs in both extracted equations because the verifier uses the same response in both. The $\mathrm{ee}$ and $\mathrm{pp}$ relations extract openings of $\cc{e}{i}$ and $\cc{p}{i}$. The $\mathrm{ep}$ relation extracts a second opening of the anchor $\cc{p}{i}$; its message must agree with the $\mathrm{pp}$ message unless binding is broken. Its base is the already opened $\cc{e}{i}$. Two efficiently obtained openings $(m,\rho),(m',\rho')$ with $m\ne m'$ yield $\log_g h=(m-m')/(\rho'-\rho)$, so binding ensures computational consistency of the extracted messages. Substituting the known base opening gives
\begin{equation}\label{eq:fork-product-opening}
\cc{\lrel}{i}=g^{m^{\bas(\lrel)}_i m^{\anc(\lrel)}_i}\,h^{\rho^{\bas(\lrel)}_i m^{\anc(\lrel)}_i+u^{\lrel}_i},
\end{equation}
Thus the product commitments open to $(\mc{e}{i})^2$, $(\mc{p}{i})^2$, and $\mc{e}{i}\mc{p}{i}$, and homomorphic reconstruction gives an opening of $c'_{\dist}$ to $\dist(\vm{e},\vm{p})\bmod\ord$. The assumed compatible threshold extractor provides an opening $(\dist',\blind')$ of this same commitment with $0\le\dist'\le\thr$. If its message differs from the reconstructed residue, these two openings break binding. Define $\delta_{\mathsf{bind}}$ for any such conflict, including the $\ndim$ anchor comparisons, in this joint experiment; define $\delta_{\mathsf{G16}}$ for failure to obtain the first branch's threshold witness without changing the fork state. Removing these bad events gives the stated bound without assuming independence. No coordinate range is extracted, so the result concerns $\Rel^{\mathrm{mod}}_{\thr}$. With polynomial $Q$ and negligible bad-event probabilities, any non-negligible $p$ gives non-negligible extraction probability.

{\emph{Offline template privacy.} Consider the joint honest enrollment and accepting-session view of Definition~\ref{def:priv}, with $0\le\thr<\ord$. The simulator generates the simulated Groth16 CRS and samples the enrollment ledger vector uniformly in $\Gp^{\ndim}$. Independent Pedersen blindings make the real ledger and the $4\ndim$ authentication commitments jointly uniform, independent of the templates. The simulator samples $\vc{p},\vc{ee},\vc{pp}$ uniformly, sets $c'_{\dist}=\Com_{\ck}(0;\rho)$ for uniform $\rho$, samples the first $\ndim-1$ entries of $\vc{ep}$, and solves reconstruction for its last entry. Fixing the other $3\ndim-1$ product commitments, the map $c'_{\dist}\mapsto\cc{ep}{\ndim}$ is a bijection because $-2$ is invertible modulo the odd prime $\ord$. Consequently the full product tuple is jointly uniform; $c'_{\dist}$ has the correct marginal and remains the prescribed function of that tuple, not an extra independent value. The simulator knows the valid threshold witness $(0,\rho)$.

It samples uniform $\chal,\bm z$, reconstructs
\begin{equation}\label{eq:sigma-simulated-first-messages}
\begin{aligned}
\alpha_i^{\lrel}&=g^{z_i^{\lrel,\mathrm m}}h^{z_i^{\lrel,\mathrm r}}(\cc{\anc(\lrel)}{i})^{-\chal},\\
\beta_i^{\lrel}&=(\cc{\bas(\lrel)}{i})^{z_i^{\lrel,\mathrm m}}h^{z_i^{\lrel,\mathrm u}}(\cc{\lrel}{i})^{-\chal},
\end{aligned}
\end{equation}
and programs the canonically encoded query in \eqref{eq:chal} to $\chal$. The observer may query the oracle before and after receiving the view. Programming aborts only if this input was already queried. A fresh first-message component is uniform even conditional on the commitments and prior oracle view, giving the conservative bound $q_{\Hash}/\ord$ on that event. Later queries use the programmed table.

For the hybrid justification, first replace the honest Groth16 CRS/proof by their simulated counterparts for the same true statement. Next replace the honest Sigma transcript by its programmed simulation, using the bijection between independent masks and responses for fixed challenge. The resulting view depends on the templates only through the uniformly distributed commitments, so replace their generation by the bijective construction above and recompute the simulated proof for the resulting true threshold statement. Every $c'_{\dist}\in\Gp$ has a zero-message opening; the construction explicitly knows one. The total distinguishing bound is $\AdvSym^{\mathsf{zk}}+q_{\Hash}/\ord$. This proves the stated single honest accepting-session privacy, not privacy of arbitrary rejected transactions or a conditional simulation for an externally fixed enrollment record.}

{
\paragraph*{Proof of Proposition~\ref{prop:circ}}
\emph{(i)} Fix the contextual public parameters $(\ndim,\pol,\chi,\thr)$ with $0\le\thr<2^{48}$ and $\ndim(2^{16}-1)^2<2^{48}$. The circuit proves that the coordinates of $\vm{e}$ and $\vm{p}$ lie in $[0,2^{16})$, that $\gamma^{\mathrm{e}}=\PCom_{\mathrm{e}}(\vm{e};\blind^{\mathrm{e}};\pol,0,\ndim)$ and $\gamma^{\mathrm{p}}=\PCom_{\mathrm{p}}(\vm{p};\blind^{\mathrm{p}};\pol,\chi,\ndim)$, and that the distance computed in $\Fr$ is at most $\thr$. The contract fixes $\gamma^{\mathrm{e}}$ from the ledger, fixes $(\pol,\thr)$ at deployment, and recomputes $\chi$ from $(\ctx,\id,\nonce)$; $\gamma^{\mathrm{p}}$ is the remaining public input supplied by the transaction. If $\Verify$ accepts with probability $p$, Groth16 knowledge soundness yields a satisfying witness except with probability $\AdvSym^{\mathsf{G16}}$. Part~(iii) identifies the field computation with the integer distance, so the extracted witness lies in $\Rel_{\thr,\pol,\chi}$.

\emph{(ii)} Let $\mathsf{Pack}_{\ndim}$ split a vector into blocks of fifteen coordinates and encode each block by
\begin{equation}\label{eq:coordinate-packing}
\mathsf{PackBlock}(m_0,\ldots,m_{14})=\sum_{j=0}^{14}m_j2^{16j},
\end{equation}
with deterministic zero padding in the last block and to an even number of blocks. The circuit first proves every $m_j\in[0,2^{16})$. The encoding in \eqref{eq:coordinate-packing} therefore stays below $2^{240}<\fieldord$, so no field reduction occurs and, for fixed $\ndim$, $\mathsf{Pack}_{\ndim}$ is injective; absorbing $\ndim$ also separates encodings whose zero padding could otherwise look alike. The complete enrollment encoding is
\begin{equation}\label{eq:poseidon-enrollment-encoding}
(\mathrm{DST}_{\mathrm e},\mathrm{FORMAT},\pol,0,\ndim,\blind^{\mathrm e},\mathsf{Pack}_{\ndim}(\vm{e})),
\end{equation}
and the probe encoding replaces the role tag, zero context, salt, and vector by $(\mathrm{DST}_{\mathrm p},\chi,\blind^{\mathrm p},\vm p)$. Suppose two distinct complete encodings have the same final chain value. If their initialization tuples differ but their initial states agree, they collide in the six-input initialization compression. Otherwise follow the two chains from their first differing state or packed-block pair to the first later state at which they agree; the two distinct three-input compression inputs at that step have the same output. Thus a collision of the commitment yields a collision in one of the assumed Poseidon-chain compression functions. Conditional on honest enrollment, the enrolled opening and the extracted opening both map to the ledger value $\gamma^{\mathrm e}$; except with probability $\AdvSym^{\mathsf{cr}}$, their complete encodings, and hence their vectors, are equal. No such ``original enrolled vector'' conclusion is asserted for an arbitrary ledger element not produced by honest $\Enroll$. Collision resistance here is an explicit assumption on this fixed-format compression family, not a reduction to a standard hardness problem or a collision in the full Poseidon permutation.

\emph{(iii)} Each coordinate difference satisfies $|\mc{e}{i}-\mc{p}{i}|\le2^{16}-1$. Therefore the integer sum is at most $\ndim(2^{16}-1)^2<2^{48}<\fieldord$. The coordinate range constraints make every field square the reduction of the corresponding integer square, and the bound prevents the accumulated sum from wrapping in $\Fr$. The circuit range-checks both the sum and $\thr$ to $48$ bits before applying the comparator, so its field constraint is exactly the integer inequality \eqref{eq:dist} for the stated profile.

\emph{(iv)} Here $\AdvSym^{\mathsf{hide}}$ denotes the standard IND hiding advantage between commitments to two chosen equal-length messages under a fresh uniform salt, for each fixed role and context. The simulator first runs the Groth16 $\mathsf{SimSetup}$, publishes its simulated CRS, and retains the simulation state. Given only $\Leak=(\mathsf{accept},\ndim,\thr,\pol,\id,\nonce,\ctx)$, it computes $\chi$, sets the dummy vectors $\vm{e}_0=\vm{p}_0=\bm 0$, samples independent uniform salts, computes the corresponding role- and context-separated commitments $(\gamma^{\mathrm e}_0,\gamma^{\mathrm p}_0)$, and invokes $\mathsf{SimProve}$ on the public statement $(\gamma^{\mathrm e}_0,\gamma^{\mathrm p}_0,\pol,\chi,\thr)$. This statement is true because the dummy distance is zero and $\thr\ge0$. The simulator outputs the joint view consisting of the simulated CRS, the dummy enrollment ledger entry, and this accepting session.

For the hybrid argument, $H_0$ is the honest joint view. In $H_1$, Groth16 zero knowledge replaces the honest CRS-and-proof view by the output of $\mathsf{SimSetup}$ and $\mathsf{SimProve}$ for the same real public statement; this endpoint statement is true, and the gap is at most $\AdvSym^{\mathsf{zk}}$. Let $X_{\mathrm{real}}$ be the pair of real enrollment and probe commitments in $H_1$, and let $X_0$ be the dummy pair above. A two-commitment hybrid using independent salts gives $X_{\mathrm{real}}\mathrel{\approx_c}X_0$ with advantage at most $2\AdvSym^{\mathsf{hide}}$. The simulated CRS and its retained state are generated independently of those salts. Fix a total PPT extension of $\mathsf{SimProve}$ on all public inputs; applying this PPT map preserves the indistinguishability of the two commitment-pair distributions. The zero-knowledge guarantee is invoked only at the true $H_0$--$H_1$ endpoint, not for any false intermediate statement used to prove the commitment-pair hybrid, and both $X_{\mathrm{real}}$ and $X_0$ correspond to true statements. Replacing $X_{\mathrm{real}}$ by $X_0$ in $H_1$ therefore yields exactly the simulator's output, with total advantage at most $\AdvSym^{\mathsf{zk}}+2\AdvSym^{\mathsf{hide}}$. This proves only the honest-enrollment, one-honest-accepting-session joint-view definition: it gives neither privacy conditioned on a fixed external ledger commitment nor privacy for arbitrary rejected or malicious transactions. The required total simulator is supplied by the trapdoor-based construction of Groth~\cite[Secs.~2.2 and~3]{groth2016size}: it samples two proof exponents and solves the linear verification equation for the third, using the nonzero setup denominator. This computation is defined for every well-formed public field vector. Its zero-knowledge guarantee is used here only for true statements.
}

\paragraph*{Proof of Proposition~\ref{prop:leak}}
If $z_i=b+\chal m_i$ with common $b$, then $z_i-z_j=\chal(m_i-m_j)$ in $\Zq$. Conditioned on $\chal\ne0$, division by the public challenge recovers the difference; the excluded zero challenge has probability $1/\ord$ for an honest uniform challenge. For a fixed challenge independent of fresh masks, each map $b_i\mapsto z_i$ is a translation of $\Zq$, hence the response vector is uniform. A fresh ROM answer gives the same challenge independence; the complete Fiat--Shamir view still requires the simulator and prior-query bound above. The synthetic experiment of \S\ref{sec:security:aux} (fixed seed, $1000$ trials) is in \texttt{analysis/verify\_claims.py}.

\section{Background}\label{app:background}

\paragraph*{Blockchain and smart contracts}
A blockchain records transactions under a consensus protocol; smart-contract platforms also execute contracts in a deterministic virtual machine~\cite{nakamoto2008bitcoin,buterin2014next}. In our protocols, the contract stores enrollment commitments, enforces per-identity nonce freshness, verifies Groth16 proofs, and updates the nonce on acceptance. For \Full and \Spot, it also checks selected coordinate equations and reconstructs the distance commitment.

\paragraph*{Homomorphic operations on Pedersen commitments}
For $c=\Com_{\ck}(m;\blind)=g^{m}h^{\blind}$, addition and subtraction of committed values follow from
\begin{equation}\label{eq:pedersen-homomorphism}
\begin{aligned}
\Com_{\ck}(m;\blind)\,\Com_{\ck}(m';\blind')&=\Com_{\ck}(m{+}m';\blind{+}\blind'),\\
\Com_{\ck}(m;\blind)\,\Com_{\ck}(m';\blind')^{-1}&=\Com_{\ck}(m{-}m';\blind{-}\blind').
\end{aligned}
\end{equation}
To verify a product, we link its commitment to the factor commitments using a Sigma protocol made non-interactive by the Fiat--Shamir transform: fresh first messages are hashed with the statement and context to derive the challenge, and affine responses are checked through commitment equalities. Section~\ref{sec:strategies:hom} uses three instances per coordinate: $\mathrm{ee}$, $\mathrm{pp}$, and $\mathrm{ep}$.

\paragraph*{Zero-knowledge proofs}
A non-interactive proof system consists of key generation, proof generation, and verification. The homomorphic strategies use Groth16~\cite{sasson2014zerocash,groth2016size} to prove the opening-and-range relation
\begin{equation}\label{eq:homomorphic-threshold-relation}
\bigl\{\bigl((c'_{\dist},\thr),(\dist,\blind_{\dist})\bigr):\ c'_{\dist}=\Com_{\ck}(\dist;\blind_{\dist})\ \wedge\ 0\le\dist\le\thr\bigr\}.
\end{equation}
Equation~\eqref{eq:homomorphic-threshold-relation} captures only the threshold statement. The proof is compact, and its verification cost is independent of the biometric dimension; \Full and \Spot check coordinate-wise consistency outside this fixed circuit. \Circ instead proves the contextual relation in \eqref{eq:context-relation}, including vector commitments, coordinate ranges, and the integer distance threshold, in one circuit.

\section{Provenance of the Reported Numbers}\label{app:prov}

{Table~\ref{tab:provenance} maps the reported measurements to files in the public \href{https://github.com/xiaolin9595/BIOZERO-Verifiable-On-Chain-Biometric-Authentication-}{\sysname{} repository}. All paths are relative to the repository root and independent of the benchmark host's directory structure.}

\begin{table*}[t]
\caption{Sources of the reported measurements.}
\label{tab:provenance}
\centering\footnotesize
\begin{tabular}{@{}p{4.0cm}p{13.2cm}@{}}
\toprule
Quantity & Source \\
\midrule
CircuitCheck grid and combined Figure~\ref{fig:isc} data &
\nolinkurl{paper/data/circuitcheck-grid-enine-2026-10-28.json} (SHA-256 \texttt{b738134e\ldots{}a5d3d}); \nolinkurl{paper/data/fig3-with-circuitcheck.csv} (\texttt{64697fbd\ldots{}e5b1a}); PASS report \nolinkurl{paper/data/fig3-with-circuitcheck-verification.json} (\texttt{ffec9602\ldots{}ee40}) \\
Current \Full stand-in transaction on enine &
\nolinkurl{paper/data/hom-complete-enine-2026-10-28.json} (SHA-256 \texttt{f4e9a6f0\ldots{}0b71a}); PASS report \nolinkurl{paper/data/hom-complete-enine-2026-10-28-verification-spotcheck.json} (\texttt{55f13add\ldots{}16d34}) \\
Current \Spot stand-in dimension sweep: $t$, gas, calldata, and controls &
\nolinkurl{paper/data/spotcheck-n-sweep-enine-2026-10-28.json} (SHA-256 \texttt{514f56e3\ldots{}94a11}) and \nolinkurl{paper/data/spotcheck-n-sweep-enine-2026-10-28.csv} (\texttt{0615c054\ldots{}17b19}) \\
\Spot sweep verification and manifest &
\nolinkurl{paper/data/spotcheck-n-sweep-enine-2026-10-28-verification.json} (SHA-256 \texttt{b154ab5c\ldots{}36b796}); \nolinkurl{paper/data/spotcheck-n-sweep-enine-2026-10-28-manifest.json} (\texttt{97a1a39c\ldots{}ef332}) \\
Current complete \Circ transaction: gas, calldata, enrollment, controls &
\nolinkurl{paper/data/circuit-complete-enine-2026-10-28.json} (SHA-256 \texttt{f01811a1\ldots{}ca0d}); CSV summary \nolinkurl{paper/data/complete-gas-summary-enine-2026-10-28.csv} (\texttt{5d619dd5\ldots{}945bc}) \\
Homomorphic kernel gas, detection of one bad coordinate, unsampled tamper accepted & \texttt{paper/data/pedersen-bn254-gas.json} (SHA-256 \texttt{751d7182\ldots{}35cf3}), recorded 2026-09-08 \\
Bare Groth16 verifier gas of \Circ & \texttt{paper/data/routea-verifier-gas.json} (SHA-256 \texttt{54d290ed\ldots{}06e8}), recorded 2026-09-08 \\
Current packed \Circ constraints, witness/proving time, and process maximum RSS &
\nolinkurl{paper/data/packed-proving-enine-2026-10-28.json} (SHA-256 \texttt{a0d475a5\ldots{}7686}); raw CSV \nolinkurl{paper/data/packed-proving-enine-2026-10-28-raw.csv} (\texttt{4fb32f85\ldots{}e6f6}) and summary CSV \nolinkurl{paper/data/packed-proving-enine-2026-10-28-summary.csv} (\texttt{d574380c\ldots{}2147}); five cold processes per dimension \\
Later enine \Circ campaign: environment, setup, and verification &
\nolinkurl{paper/data/remaining-enine-2026-10-28-environment.json} (\texttt{9d7fab9a\ldots{}c4a1}); \nolinkurl{paper/data/remaining-enine-2026-10-28-setup-summary.json} (\texttt{ea07ed87\ldots{}1fb0}); PASS \nolinkurl{paper/data/remaining-enine-2026-10-28-verification.json} (\texttt{e4a76289\ldots{}2ed2}) \\
Tables and figures & generated by \texttt{paper/analysis/make\_tables.py}, \texttt{paper/analysis/make\_figures.py}, and \texttt{paper/analysis/make\_fig3\_comparison.py} from the recorded data \\
Difference-leak, group-parameter, and single-coordinate checks & \texttt{paper/analysis/verify\_claims.py} (fixed seed; synthetic data; example modulus and ranges) \\
{Vanilla and threshold-circuit baseline series in Figure~\ref{fig:isc}: proving, verifier and deployment gas, serialized size, and verifier creation bytecode (runtime separately for EIP-170)} &
{\nolinkurl{paper/data/fresh-native-fig3-full-2026-10-28.json} (SHA-256 \texttt{603366ed\ldots{}a996}) and \nolinkurl{paper/data/fresh-native-fig3-full-2026-10-28.csv} (\texttt{5b643b1f\ldots{}fd95}); PASS report \nolinkurl{paper/data/fresh-native-fig3-full-2026-10-28-verification.json} (\texttt{6e238e44\ldots{}7e60})} \\
Quantization table &
FEI: \nolinkurl{paper/data/fei-mobilefacenet128-shared512.json} (\texttt{97d217d6\ldots{}a6d72c}) and \nolinkurl{paper/data/fei-inceptionresnetv1-vggface2-512.json} (\texttt{6eaa9075\ldots{}90e003}); LFW: \nolinkurl{paper/data/lfw-mobilefacenet128-harmonized.json} (\texttt{6264e576\ldots{}fae83}) and \nolinkurl{paper/data/lfw-inceptionresnetv1-vggface2-harmonized.json} (\texttt{a8645d08\ldots{}6f6b8}). Each LFW aggregate preserves its source-result and independent-verification SHA-256; \nolinkurl{paper/analysis/make_tables.py} and \nolinkurl{paper/analysis/verify_claims.py} independently validate the fixed-threshold rows. \\
Reconstructed unpacked-ablation constraint check &
\nolinkurl{paper/data/reconstructed-unpacked-ablation-enine-2026-10-28.json} (SHA-256 \texttt{49345fb4\ldots{}ad7d}); CSV \nolinkurl{paper/data/reconstructed-unpacked-ablation-enine-2026-10-28.csv} (\texttt{079b837d\ldots{}6755}) \\
{Recovered Figure~\ref{fig:isc} campaign, historical comparison only} &
{\nolinkurl{paper/data/recovered-native-fig3-full-2026-10-28.json} (SHA-256 \texttt{6bf110a5\ldots{}5add}) and PASS report \nolinkurl{paper/data/recovered-native-fig3-full-verification-2026-10-28.json} (\texttt{dfaa1a9b\ldots{}5712}); no recovered sample enters a fresh aggregate} \\
\bottomrule
\end{tabular}
\end{table*}

In the kernel file, the flag \texttt{unsampledTargetTamperAccepted} is true for every sampled configuration and false for the exhaustive check, and \texttt{oneBadCoordinateDetection} equals $\nsamp/\ndim$ for each configuration, as Proposition~\ref{prop:accept} predicts.

The experiments ran on \texttt{enine}, with an Intel Xeon w7-3445, 251.36\,GiB RAM, Ubuntu~22.04.5, and Linux~6.8.0-138. The complete-\Circ, packed, and reconstructed-unpacked experiments used Node.js~22.22.3, Circom~2.2.3, snarkjs~0.7.6, \texttt{solc}~0.8.36, Ganache~7.9.2, and ethers~5.8.0. The seven-size CircuitCheck grid used \texttt{solc}~0.8.33 and ethers~6.16.0. Complete-transaction costs are medians of three transactions; packed-circuit timings and process maximum RSS are medians of five cold processes per dimension. FEI and LFW inference used the recorded CPU providers.

{The \Full benchmark used a fresh single-party threshold setup. The \Spot sweep reused that setup and generated three authentications per $(N,\eta)$ setting, each with a fresh transcript and proof. The 28 labelled settings represent 22 distinct $(N,t)$ pairs because $t=\min\{N,\lceil\eta\log_2N\rceil\}$ saturates at small dimensions. The sweep passes all 294 verification checks. The implementation uses $h=5G$ and a threshold circuit that checks $\dist\le\thr$ without proving the Pedersen opening; its timings measure this simplified circuit.}

{The Vanilla and threshold-circuit baseline series in Figure~\ref{fig:isc} use the fresh 2026-10-28 measurements listed in the table. Environment and source-transform manifests record the toolchain, source repair, and reused setup artifacts. Vanilla used default Rayon settings for $N\le32$ and one thread for $N=64,128$; the threshold-circuit baseline also used one thread. Proving aggregates are five-run means with sample standard deviations after one warm-up. Gas means use 30 threshold-circuit baseline transactions after five warm-ups and five Vanilla transactions per dimension after one warm-up. These gas transactions reuse a cached proof. The CircuitCheck grid uses ten timed processes and five distinct-proof verifier transactions per dimension, after one warm-up, with CPU~0 affinity and single-thread environment settings.}

{Vanilla and the threshold-circuit baseline reuse their compiled circuits, proving keys, and matching verifiers; CircuitCheck uses fresh benchmark keys. The Vanilla repair removes a local all-zero \texttt{res} declaration that shadows the public argument. The threshold-circuit baseline evaluates a dimension-independent threshold/opening component, identified by the target \texttt{biozero-fixed} in the source records. At $N=128$, Vanilla's 34,837-byte runtime exceeds EIP-170. The standard-limit failure and unlimited-local-chain measurements are recorded in \nolinkurl{paper/data/fresh-native-fig3-full-n128-chain-controls-2026-10-28.json}.}

{The recovered 2026-06-05 campaign is retained in \nolinkurl{paper/data/recovered-native-fig3-full-2026-10-28.json} for historical comparison only; its samples are excluded from the reported 2026-10-28 aggregates.}

\bibliographystyle{IEEEtran}
\bibliography{refs}

\end{document}